\documentclass[runningheads]{llncs}
\usepackage[T1]{fontenc}
\usepackage{graphicx}
\usepackage{amsmath,amsfonts,amssymb}%amssymb
\usepackage{graphicx}
\usepackage{textcomp}
\usepackage{xcolor}
\def\BibTeX{{\rm B\kern-.05em{\sc i\kern-.025em b}\kern-.08em
		T\kern-.1667em\lower.7ex\hbox{E}\kern-.125emX}}
\usepackage{etoolbox}
\usepackage{multirow}
\usepackage{tabularx}

\newcolumntype{C}{>{\centering\arraybackslash}X}
\usepackage{booktabs}
\usepackage{lipsum}
\usepackage{subcaption}
\usepackage{pgfplots}
\pgfplotsset{compat=1.16}
\newtheorem{lem}{Lemma}
\usepackage{algorithmic,float}
\usepackage[ruled,linesnumbered,noend]{algorithm2e}%
\let\oldnl\nl% Store \nl in \oldnl
\newcommand{\nonl}{\renewcommand{\nl}{\let\nl\oldnl}}% Remove line number for one line
\usepackage{caption} %using comment \captionsetup in Figure 6
\makeatletter
\patchcmd\algocf@Vline{\vrule}{\vrule \kern-0.4pt}{}{}
\patchcmd\algocf@Vsline{\vrule}{\vrule \kern-0.4pt}{}{}
\makeatother
\SetKwInput{KwInput}{Input}                % Set the Input
\SetKwInput{KwOutput}{Output}              % set the Output
\usepackage{tikz}
\usepackage{environ}
\newcommand\scalemath[2]{\scalebox{#1}{\mbox{\ensuremath{\displaystyle #2}}}}%\scalemath in Def 3.17

\definecolor{ballblue}{rgb}{0.13, 0.67, 0.8}

\newlength\myindent
\makeatletter
\newcommand{\labeltext}[2]{%
	\@bsphack
	\csname phantomsection\endcsname % in case hyperref is used
	\def\@currentlabel{#1}{\label{#2}}%
	\@esphack
}
\makeatother
\newcommand*{\thead}[1]{\multicolumn{1}{|c|}{\bfseries #1}}

\usetikzlibrary{shapes.geometric,arrows,arrows.meta,calc,fit,matrix}
\definecolor{arrowcolor}{rgb}{.27,.45,.77}
\tikzstyle{arrow} = [arrowcolor,opacity=1,thin, {Triangle[angle=60:1.2mm]}-{Triangle[angle=60:1.2mm]}]
\tikzstyle{chart} = [rectangle, minimum width=3cm, minimum height=1cm, text centered,text width =3cm, draw=black, fill=white!30]
\tikzstyle{circlular} = [circle, minimum width=1cm, minimum height=1cm, text centered,text width =1cm, draw=black, fill=white!30]
\tikzstyle{circ} = [circle, minimum width=5mm, minimum height=5mm, text centered,text width =5mm, draw=black, fill=yellow!30]
\tikzstyle{roundrect3} = [rectangle,rounded corners, minimum width=3cm, minimum height=1.6cm, text centered,text width =1.4cm, draw=black, fill=red, text opacity=1]
\tikzstyle{roundrect4} = [rectangle,rounded corners, minimum width=3cm, minimum height=1.6cm, text centered,text width =1.4cm, draw=black, fill=blue, text opacity=1]
\tikzstyle{roundrect6} = [rectangle,rounded corners, minimum width=3cm, minimum height=1.6cm, text centered,text width =1.4cm, draw=black, fill=green, text opacity=1]
\tikzstyle{invisible} = [minimum width=3cm, minimum height=2cm, text centered,text width =1.4cm, draw=black, opacity = 1]
\tikzstyle{tinyarrow} = [thin,->,>=stealth]
\tikzstyle{strippedline} = [thick,dotted,>=stealth]
\BeforeBeginEnvironment{appendices}{\clearpage}

\makeatletter
\newcommand{\removelatexerror}{\let\@latex@error\@gobble}
\makeatother
\begin{document}
\title{Distributed Seasonal Temporal Pattern Mining}
\author{Van Ho-Long\inst{1,2} \and
Nguyen Ho\inst{3} \and
Anh-Vu Dinh-Duc\inst{1,2} \and
Ha Manh Tran\inst{2} \and
Ky Trung Nguyen\inst{1,2} \and
Tran Dung Pham\inst{4} \and
Quoc Viet Hung Nguyen\inst{4}
}
\authorrunning{F. Author et al.}
% First names are abbreviated in the running head.ddav
\institute{School of Computer Science and Engineering, International University,Ho Chi Minh City, Vietnam\\
\email{\{hlvan,ddavu,ntky\}@hcmiu.edu.vn}\\	 \and
Vietnam National University, Ho Chi Minh City, Vietnam \\
\email{hatm@vnuhcm.edu.vn}\\ \and
Department of Computer Science, Loyola University Maryland, Baltimore, USA
\email{tnho@loyola.edu}\\ \and
Griffith University, Australia \\
\email{henry.nguyen@griffith.edu.au}\\
}
\maketitle              % typeset the header of the contribution
\begin{abstract}
The explosive growth of IoT-enabled sensors is producing enormous amounts of time series data across many domains, offering valuable opportunities to extract insights through temporal pattern mining. Among these patterns, an important class exhibits periodic occurrences, referred to as \textit{seasonal temporal patterns} (STPs). However, mining STPs poses challenges, as traditional measures such as support and confidence cannot capture seasonality, and the lack of the anti-monotonicity property results in an exponentially large search space. Existing STP mining methods operate sequentially and therefore do not scale to large datasets.
In this paper, we propose the Distributed Seasonal Temporal Pattern Mining (DSTPM), the first distributed framework for mining seasonal temporal patterns from time series. DSTPM leverages efficient data structures, specifically distributed hierarchical lookup hash structures, to enable efficient computation. Extensive experimental evaluations demonstrate that DSTPM significantly outperforms sequential baselines in runtime and memory usage, while scaling effectively to very large datasets.
\end{abstract}
%
%Experiment
%
\vspace{-0.07in}
\section{Introduction}\vspace{-0.01in}
The growing adoption of IoT systems enables the large-scale collection of time series data across domains such as energy, climate, and healthcare. Analyzing these time series can uncover hidden patterns and yield valuable insights that support decision-making and planning. Commonly, pattern mining techniques such as sequential pattern mining (SPM) \cite{sequential_lam2014mining,sequential_huang2008general} and temporal pattern mining (TPM) \cite{ho2020efficient,lee2020z} are employed to identify frequent (temporal) relationships among events. In SPM, events are analyzed in sequential order, whereas TPM incorporates additional temporal attributes such as event occurrence times, resulting in more expressive and comprehensive representations of event relationships.
A particularly useful category of temporal patterns observed in numerous real-world applications comprises those that exhibit periodic occurrences. These patterns manifest as concentrated occurrences within specific time intervals and recur at regular intervals over time. Consequently, they are referred to as \textit{seasonal temporal patterns}. %Here, the term \textit{seasonal} indicates the periodic re-occurrence, while the term \textit{temporal pattern} indicates patterns that are formed by the temporal relations between events, such as follows, contains, overlaps. Seasonal temporal patterns are useful in revealing seasonal information of temporal events and their relations. 
For instance, in the healthcare domain, experts may be interested in identifying seasonal diseases within a specific geographical region.
An illustrative example is a seasonal temporal pattern involving weather and epidemic events observed in Kawasaki, Japan \cite{diseasedata}: \{High Humidity \textit{overlaps} Low Temperature \textit{followed by} High Influenza Cases\}. This pattern recurs annually and is concentrated in January and February. Detecting such seasonal diseases can assist health experts in prevention and strategic planning.
Similarly, in the marketing domain, identifying the sequential occurrence of search keywords that appear seasonally in search engines can provide valuable insights into customer behavior, thereby supporting the development of more effective marketing strategies.

%For instance, from Fig. 1, there is a relationship between temperature, humidity, and influenza disease, indicating that the pattern “Very Low Temperature $\succcurlyeq$ Very High Humidity (meaning Very Low Temperature contains Very High Humidity), Very Low Temperature $\succcurlyeq$ High Influenza, and Very High Humidity $\succcurlyeq$ High Influenza” only occurs in January, February, and March. Detecting such seasonal temporal patterns supports the government in diminishing influenza disease annually. Another example is in the electricity supply chain in Great Britain \cite{thornton2017relationship}. The seasonal temporal patterns can reveal the relationship between the electricity demand and the renewable supply. For example, the pattern “Strong Wind $\succcurlyeq$ Much Wind Power Generation“ (meaning Strong Wind contains Much Wind Power Generation) occurs in winter, showing that there is the availability of wind power during winter. \cite{thornton2017relationship} also shows that the electricity demand tends to be highest in winter since there are fewer daylight hours and colder temperatures. i.e., more lighting and more heating during the winter months. These detections indicate that wind power can contribute to the supply mix during high demand in winter, and a spread of wind turbines across Great Britain would maximize the average availability of wind power during high demand.

\textbf{Challenges.} Although seasonal temporal patterns are highly valuable, mining them presents several challenges. First, the traditional \textit{support} measure used in pattern mining is insufficient for capturing seasonal patterns, as it does not account for their seasonality characteristics. Second, the complex relationships among events in temporal patterns lead to an exponentially large search space. Finally, seasonal temporal patterns do not satisfy the anti-monotonicity property; that is, non-empty subsets of a seasonal temporal pattern may not themselves be seasonal, mining seasonal temporal patterns is more computationally expensive as typical pruning techniques based on anti-monotonicity cannot be applied. These challenges underscore the need for efficient seasonal temporal pattern mining approaches with effective pruning strategies to manage the exponential search space. Existing approaches such as  \cite{kiran2015discovering,periodic_kiran2019discovering} focus on mining seasonal itemsets sequentially, and thus, do not scale well to large datasets. 

\textbf{Contributions.} Our key contributions are as follows. (1) Based on our sequential seasonal temporal pattern mining \cite{hoseasonal2023mining}, we propose the first distributed algorithm for seasonal temporal pattern mining (DSTPM) to support large-scale datasets. DSTPM distributes the computation across a computing cluster, thereby reducing memory usage and computational cost on each machine. (2) DSTPM employs efficient data structures, distributed hierarchical lookup hash structures, to operate in a distributed manner. Moreover, we use a measure, \textit{maxSeason}, which upholds the anti-monotonicity property, enabling efficient pruning of infrequent seasonal patterns. (3) We conduct extensive experimental evaluations on real-world and synthetic datasets, demonstrating that DSTPM outperforms the sequential baseline and scales well to very large datasets. 

%\vspace{-0.05in} 
\section{Related work}\label{sec:relatedwork}\vspace{-0.02in} 
%\textit{Seasonal pattern mining (SPM).} 
Identifying seasonal patterns that capture temporal periodicity in time series has long been a key research focus, inspiring various techniques for discovering periodic subsequences. Such techniques, initially introduced by Han et al. \cite{han1998mining,han1999efficient} and subsequently extended in later studies \cite{assfalg2009periodic,motif_zhang2022efficient,motif_liu2016fast,motif_mohammad2013approximately}, are collectively known as motif discovery techniques. However, because motifs are defined as similar time series subsequences, motif discovery focuses on identifying recurrent subsequences rather than periodic temporal patterns. 

Another prominent research direction focuses on periodic association rules \cite{tanbeer2009discovering,kiran2015discovering,amphawan2009mining,periodic_fournier2022tspin,periodic_kiran2020discovering,periodic_kiran2019discovering}, which aim to uncover seasonal associations among itemsets—for instance, market-basket analysis revealing the seasonal relationship {Glove $\Rightarrow$ Winter Hat} during the winter season. To mine such seasonal itemset patterns in transactional databases, Tanbeer et al. \cite{tanbeer2009discovering} proposed the PFP-growth algorithm, which employs \textit{minSup} and \textit{maxPer} as measures of seasonality. In this method, \textit{maxPer} enforces the periodic constraint and \textit{minSup} controls the frequency of pattern occurrences. Although PFP-growth captures seasonality through the \textit{maxPer} measure, its reliance on the \textit{minSup} threshold prevents it from identifying rare seasonal patterns. 
In a more recent study, Uday et al. \cite{kiran2015discovering} proposed the RP-growth algorithm to identify recurring itemset patterns in transactional databases. RP-growth employs an RP-tree to maintain frequent itemsets and recursively mines this structure to discover recurring ones. In their subsequent works, the authors introduced several enhancements to \cite{kiran2015discovering}. In \cite{kiran2016efficient}, they presented the Periodic-Frequent Pattern-growth++ (PFP-growth++) algorithm, which incorporates two new concepts—local periodicity and periodicity—to capture locally and globally optimal recurring patterns. This design enables a two-phase pruning process that improves runtime efficiency. Later, in \cite{periodic_kiran2019discovering}, PFP-growth++ was extended to discover periodic spatial patterns in spatio-temporal databases, and in \cite{periodic_kiran2020discovering}, it was further extended to mine maximal periodic frequent patterns. Nevertheless, all these approaches are limited to discovering seasonal relationships among itemsets.   

Several distributed approaches have been proposed for sequential pattern mining \cite{chen2017distributed,miliaraki2013mind,tanbeer2009parallel} and temporal pattern mining \cite{hodistributed2021efficient}. However, no distributed methods have been developed for seasonal temporal pattern mining. To the best of our knowledge, the proposed DSTPM represents the first distributed approach for seasonal temporal pattern mining in the literature.
\section{Preliminaries}\label{sec:preliminary}
\subsection{Temporal Event and Temporal Relation}
\textbf{Definition 3.1} (Time granularity) Given a time domain $\mathcal{T}$, a \textit{time granularity} $G$ is a \textit{complete and non-overlapping equal partitioning} of $\mathcal{T}$. Each non-empty partition $G_i \in G$ is called a (time) \textit{granule}. The position of a granule $G_i$ in $G$, denoted as $p(G_i)$, is identified by counting the number of granules which appear before and up to (including) $G_i$. 

Consider the time domain $\mathcal{T}$ consisting of an ordered set of minutes. Here, $\mathcal{T}$ can have different time granularities such as Minute, 5-Minutes, or even Hour, Day, Year. The Minute granularity consists of a set of Minute granules, each representing one unique minute. The position of granule Minute$_2$ in the Minute granularity is $p({Minute}_2)=2$. 

\hspace{-0.2in}\textbf{Definition 3.2} (Time series) A \textit{time series} $X=  x_1, x_2, ..., x_n$ in the time domain $\mathcal{T}$ is a sequence of data values that measure the same phenomenon during an observation time period, and are chronologically ordered.  

\hspace{-0.2in}\textbf{Definition 3.3} (Temporal event in a time series) A \textit{temporal event} $E$ in a time series $X$ is a tuple $E = (\omega, T)$ where $\omega \in \Sigma_X$, and $T=\{[t_{s_i}, t_{e_i}]\}$ is the set of time intervals during which $X$ has the value $\omega$. $\Sigma_X$, called the \textit{symbol alphabet} of $X$, is the finite set of permitted symbols used to encode X. $t_{s_i}$ is the start time, and $t_{e_i}$ is the end time of each time interval.

\textbf{Instance of a temporal event:} The tuple $e = (\omega, [t_{s_i}, t_{e_i}])$ is called an \textit{instance} of the temporal event $E= (\omega, T)$, representing a single occurrence of $E$ during $[t_{s_i}, t_{e_i}]$. We use the notation $E_{\triangleright e}$ to denote that the event $E$ has an instance $e$. 

\textbf{Relations between temporal events:} 
Let $E_i$ and $E_j$ be two temporal events, and $e_i=(\omega_i,[t_{s_i}, t_{e_i}])$, $e_j=(\omega_j,[t_{s_j}, t_{e_j}])$ be their corresponding instances. We rely on the popular Allens relation model \cite{allen} to define 3 basic temporal relations: \textit{Follows} ($\rightarrow$), \textit{Contains} ($\succcurlyeq$), \textit{Overlaps} ($\between$) between $E_i$ and $E_j$.  The \textit{Follows} relation represents sequential occurrences of one event after another. On theother hand, in a \textit{Contains} relation, one event occurs entirely within the timespan of another event. Finally, in an \textit{Overlaps} relation, the timespans of the two occurrences overlap each other.
%We avoid the exact time mapping problem in Allens relations by adding a tolerance \textit{buffer} $\epsilon$ to the relation's endpoints, while ensuring the relations are \textit{mutually exclusive}. Table \ref{tbl:relations} illustrates the 3 relations and their conditions, with $\epsilon \ge 0$ representing the buffer size, and $d_o$ representing the minimal overlapping duration between two event instances in an Overlaps relation.

\hspace{-0.2in}\textbf{Definition 3.4} (Temporal pattern) Let $\Re$$=$\{Follows, Contains, Overlaps\} be the set of temporal relations. A \textit{temporal pattern} $P$ is a list of triples $(r_{\textit{ij}},E_{i},E_{j})$, each representing a relation $r_{\textit{ij}} \in \Re$ between two events $E_i$ and $E_j$.

A temporal pattern of $n$ events is called an $n$-event pattern. We use $E_i \in P$ to denote that the event $E_i$ occurs in $P$, and $P_1 \subseteq P$ to say that a pattern $P_1$ is a sub-pattern of $P$. 

%An example of temporal patterns from Fig. \ref{fig:TimeSeries} is: P = <(Overlaps, Low Temperature, High Humidity), (Follows, Low Temperature, High Influenza Cases), (Follows, High Humidity, High Influenza Cases)>. Here, P is a 3-event pattern, containing pairwise temporal relations between three events: Low Temperature, High Humidity, and High Influenza Cases.

\begin{table*}
	\begin{minipage}{1\textwidth}
		\caption{A Temporal Sequence Database $\mathcal{D}_{\text{SEQ}}$}
		\label{tbl:SequenceDatabase}
		\resizebox{\textwidth}{4.5cm}{
			\begin{tabular}{ |c|c|p{12cm}| }
				\hline  {\bfseries Granules} & {\bfseries Position} & {\bfseries \;\;\;\;\;\;\;\;\;\;\;\;\;\;\;\;\;\;\;\;\;\;\;\;\;\;\;\;\;\;\;\;\;\;\;\;Temporal sequences} \\
				\hline  
				$\textbf{\textit{G}}_1$ & 1   &  (C:1,[7:00,7:10]), (C:0,[7:10,7:15]), (D:1,[7:00,7:05]), (D:0,[7:05,7:15]), (F:0,[7:00,7:10]), (F:1,[7:10,7:15]), (M:1,[7:00,7:15], (I:1,[7:00,7:10]), (I:0,[7:10,7:15])
				\\
				\hline 
				$\textbf{\textit{G}}_2$ & 2   &  (C:1,[7:15,7:20]), (C:0,[7:20,7:30]), (D:1,[7:15,7:20]), (D:0,[7:20,7:30]), (F:0,[7:15,7:20]), (F:1,[7:20,7:30]), (M:1,[7:15,7:20]), (M:0,[7:20,7:30]), (I:1,[7:15,7:30])
				\\ 
				\hline 
				$\textbf{\textit{G}}_3$ & 3   &   (C:1,[7:30,7:40]), (C:0,[7:40,7:45]), (D:1,[7:30,7:40]), (D:0,[7:40,7:45]), (F:0,[7:30,7:40]), (F:1,[7:40,7:45]), (M:1,[7:30,7:45]), (I:1,[7:30,7:45])
				\\
				\hline 
				$\textbf{\textit{G}}_4$& 4   &  (C:0,[7:45,8:00]), (D:1,[7:45,7:55]), (D:0,[7:55,8:00]), (F:0,[7:45,7:55]), (F:1,[7:55,8:00]), (M:1,[7:45,7:55]), (M:0,[7:55,8:00]), (I:1,[7:45,7:55]), (I:0,[7:55,8:00])
				\\
				\hline
				$\textbf{\textit{G}}_5$ & 5   &  (C:0,[8:00,8:15]), (D:0,[8:00,8:15]), (F:1,[8:00,8:15]), (M:1,[8:00,8:15]),  (I:1,[8:00,8:15])
				\\
				\hline 
				$\textbf{\textit{G}}_6$ & 6   &  (C:0,[8:15,8:30]), (D:0,[8:15,8:30]), (F:0,[8:15,8:30]), (M:1,[8:15,8:30]), (I:1,[8:15,8:30])
				\\ 
				\hline 
				$\textbf{\textit{G}}_7$ & 7   &  (C:1,[8:30,8:45]), (D:1,[8:30,8:45]), (F:0,[8:30,8:45]), (M:0,[8:30,8:45]), (I:0,[8:30,8:45])
				\\
				\hline 
				$\textbf{\textit{G}}_8$ & 8   &  (C:1,[8:45,9:00]), (D:1,[8:45,9:00]), (F:0,[8:45,9:00]), (M:1,[8:45,9:00]), (I:0,[8:45,9:00])
				\\
				\hline
				$\textbf{\textit{G}}_9$ & 9   &  (C:0,[9:00,9:15]), (D:0,[9:00,9:15]), (F:1,[9:00,9:15]), (M:1,[9:00,9:15]), (I:1,[9:00,9:15])
				\\
				\hline 
				$\textbf{\textit{G}}_{10}$ & 10   &  (C:0,[9:15,9:30]), (D:0,[9:15,9:30]), (F:1,[9:15,9:30]), (M:1,[9:15,9:30]), (I:1,[9:15,9:30])
				\\ 
				\hline 
				$\textbf{\textit{G}}_{11}$ & 11   &  (C:1,[9:30,9:35]), (C:0,[9:35,9:45]), (D:1,[9:30,9:35]), (D:0,[9:35,9:45]), (F:0,[9:30,9:40]), (F:1,[9:40,9:45]), (M:1,[9:30,9:45]), (I:1,[9:30,9:45])
				\\
				\hline 
				$\textbf{\textit{G}}_{12}$ & 12   &  (C:1,[9:45,9:55]), (C:0,[9:55,10:00]), (D:1,[9:45,9:50]), (D:0,[9:50,10:00]), (F:0,[9:45,9:55]), (F:1,[9:55,10:00]), (M:0,[9:45,10:00]), (I:1,[9:45,10:00])
				\\
				\hline
				$\textbf{\textit{G}}_{13}$ & 13   &  (C:0,[10:00,10:15]), (D:1,[10:00,10:10]), (D:0,[10:10,10:15]), (F:0,[10:00,10:10]), (F:1,[10:10,10:15]), (M:1,[10:00,10:15]), (I:1,[10:00,10:15])
				\\
				\hline 
				$\textbf{\textit{G}}_{14}$ & 14   &  (C:1,[10:15,10:25]), (C:0,[10:25,10:30]), (D:1,[10:15,10:25]), (D:0,[10:25,10:30]), (F:0,[10:15,10:25]), (F:1,[10:25,10:30]), (M:0,[10:15,10:30]), (I:0,[10:15,10:30])
				\\
				\hline
			\end{tabular} 
		}
	\end{minipage}	
	\vspace{-0.1in}
\end{table*}

\subsection{Temporal Sequence Database}
\textbf{Definition 3.5} (Temporal sequence of a time series) Consider a time series $X$ of granularity $G$. A \textit{temporal sequence} $Seq_i=<e_{1},..., e_n>$ is a list of $n$ event instances.
We say that $Seq_i$ has size $n$, denoted as $\mid Seq_i\mid = n$. 

%In the previous example, we obtain the temporal sequences for the granules in $H$ as: $Seq_1$ = $<$(C:1, [1, 2]), (C:0, [3, 3])$>$, $Seq_2$ = $<$(C:1, [4, 4]), (C:0, [5, 6])$>$, $Seq_3$ = $<$(C:1, [7, 8]), (C:0, [9, 9])$>$, etc.

\hspace{-0.2in}\textbf{Definition 3.6} (Temporal sequence database) 
Given a set of time series $\mathcal{X}=\{X_1,...,X_n\}$ of granularity $G$. 
A set of temporal sequences from $\mathcal{X}$ forms a \textit{temporal sequence database} $\mathcal{D}_{\text{SEQ}}$ where each row $i$ contains a temporal sequence $S_i$  that belong to the same granule in $G$.

Table \ref{tbl:SequenceDatabase} shows an example of $\mathcal{D}_{\text{SEQ}}$ using $\Sigma_X = \{0, 1\}$. There are $5$ time series \{C, D, F, M, I\} (C: Cooker, D: Dish Washer, F: Food Processor, M: Microwave, I: Iron). Each row $i$ in Table \ref{tbl:SequenceDatabase} contains the temporal sequences of \{C, D, F, M, I\} belonging to the same granule $G_i$. For instance, row $1$ includes the sequences of the $5$ time series at the first granule $G_1$: $<$(C:1,[7:00, 7:10]), (C:0,[7:10, 7:15]), (D:1, [7:00, 7:05]), (D:0, [7:05, 7:15]), (F:0, [7:00, 7:10]), (F:1, [7:10, 7:15]), (M:1, [7:00, 7:15]), (I:1, [7:00, 7:10]), (I:0, [7:10, 7:15])$>$. 
\subsection{Frequent Seasonal Temporal Pattern}\vspace{-0.02in}
\textbf{Definition 3.7} (Support set of a temporal event) Consider a temporal sequence database $\mathcal{D}_{\text{SEQ}}$ of granularity $G$, and a temporal event $E$. The set of granules $G_i$ in $\mathcal{D}_{\text{SEQ}}$ where $E$ occurs, arranged in an increasing order, is called the \textit{support set} of event $E$ and is denoted as $\text{SUP}^E = \{G_l^E, ..., G_r^E \}$, where $1 \leq l \leq r \leq |\mathcal{D}_{\text{SEQ}}|$. The granule $G_i$ at which event $E$ occurs is denoted as $G_i^E$. The support set of a group of events, denoted as $\text{SUP}^{(E_i, ..., E_k)}$, and the support set of a temporal pattern, denoted as $\text{SUP}^P = \{G_l^P, ..., G_r^P \}$, are defined similarly to that of a temporal event.

\hspace{-0.2in}\textbf{Definition 3.8} (Near support set of a temporal pattern) 
Consider a pattern $P$ with the support set $\text{SUP}^P = \{G_l^P, ..., G_r^P \}$. Let \textit{maxPeriod} be the \textit{maximum period threshold}, representing the predefined maximal period between any two consecutive granules in $\text{SUP}^P$.  The set $\text{SUP}^P$ is called a \textit{near support set} of $P$ if $\forall (G_o^P, G_p^P) \in \text{SUP}^P$$:$ $(G_o^P \text{and } G_p^P \text{are consecutive})$ $\wedge$ $|p(G_o^P)-p(G_p^P)| \le \textit{maxPeriod}$, where $p(G_o^P)$ and $p(G_p^P)$ are the positions of $G_o^P$ and $G_p^P$ in granularity $G$. We denote the near support set of pattern $P$ as $\text{NearSUP}^P$. 

Intuitively, the near support set of $P$ is a support set where $P$'s occurrences are close in time. Moreover, $\text{NearSUP}^P$ is called a \textit{maximal near support set} if $\text{NearSUP}^P$ has no other superset beside itself which is also a near support set. The near support set of an event is defined similarly to that of a pattern.

%As an example, consider the pattern $P$ = (Contains, C:1, D:1) (or C:1 $\succcurlyeq$ D:1) in Table \ref{tbl:SequenceDatabase}, and let $\textit{maxPeriod}=2$. Here, the support set of $P$ is $\text{SUP}^{P}$ = $\{G_1, G_2, G_3, G_7, G_8, G_{11}, G_{12}, G_{14}\}$. Hence, $P$ has three maximal near support sets: $\text{NearSUP}_1^{P} = \{G_1,G_2,G_3\}$, $\text{NearSUP}_2^{P} = \{G_7,G_8\}$, and $\text{NearSUP}_3^{P} = \{G_{11},G_{12},G_{14}\}$. 
%Fig. \ref{fig:seasonalpattern} illustrates the three near support sets of $P$.

\hspace{-0.2in}\textbf{Definition 3.9} (Season of a temporal pattern) Let $\text{NearSUP}^P$ be a near support set of a pattern $P$. Then $\text{NearSUP}^P$ is called a \textit{season} of $P$ if $\textit{den}(\text{NearSUP}^P)$ $=$ $\mid$$\text{NearSUP}^P$$\mid$ $\ge \textit{minDensity}$, where $\scalemath{0.95}{\textit{den}(\text{NearSUP}^P)}$ counts the number of granules in $\text{NearSUP}^P$ called the \textit{density of}  $\scalemath{0.95}{\text{NearSUP}^P}$, and $\textit{minDensity}$ is a predefined minimum density threshold.
 
%For instance, in the previous example, we have $\textit{den}(\text{NearSUP}_1^{P}) = |\text{NearSUP}_1^{P}|=3$. Similarly, $\textit{den}(\text{NearSUP}_2^{P}) = 2$, $\textit{den}(\text{NearSUP}_3^{P}) = 3$. If the occurrences of a pattern $P$ are dense enough, the near support set becomes a season of $P$. 
Intuitively, a {\em season} of a temporal pattern is a {\em concentrated occurrence period}, separated by a long {\em gap period} of no/few occurrences, before the next season starts. The season of an event is defined similarly as for a pattern.

The \textit{distance} between two seasons $\text{NearSUP}_i^P$ = $\{G_k^P, ..., G_n^P\}$ and $\text{NearSUP}_j^P$ = $\{G_r^P, ..., G_u^P\}$ is computed as:
$\textit{dist}(\text{NearSUP}_i^P, \text{NearSUP}_j^P)$ = $\mid$$p(G_n^P)-p(G_r^P)$$\mid$.

%\begin{figure}[!t]
%	\centering
%	\includegraphics[width=1\linewidth]{figures/SeasonalPattern.pdf}
%	\vspace{-0.2in}
%	\caption{Near support sets of pattern $P$ = (C:1 $\succcurlyeq$ D:1)}
%	\label{fig:seasonalpattern}
%	\vspace{-0.14in}
% \end{figure}

\hspace{-0.2in}\textbf{Definition 3.10} (Frequent seasonal temporal pattern) 
Let $\mathcal{PS}$ = $\left\{\text{NearSUP}^{P}\right\}$ be the set of seasons of a temporal pattern $P$, and $\textit{minSeason}$ be the \textit{minimum seasonal occurrence} threshold, $\textit{distInterval}$ = $[\text{dist}_{\min}, \text{dist}_{\max}]$ be the \textit{distance interval} where $\text{dist}_{\min}$ is the minimum distance and $\text{dist}_{\max}$ is the maximum distance. A temporal pattern $P$ is called a \textit{frequent seasonal temporal pattern} iff $\textit{seasons}(P)$ $=$ $\mid$$\mathcal{PS}$$\mid$ $\ge$ $\textit{minSeason}$ $\wedge$ $\forall (\text{NearSUP}_i^P$, $\text{NearSUP}_j^P)$ $\in$ $\mathcal{PS}$: they are consecutive and $\text{dist}_{\min} \leq \textit{dist}(\text{NearSUP}_i^P, \text{NearSUP}_j^P) \leq \text{dist}_{\max}$. 

Intuitively, a pattern $P$ is \textit{seasonal} if the distance between two consecutive seasons is within the predefined distance interval. Moreover, a seasonal temporal pattern is \textit{frequent} if it occurs more often than a predefined \textit{minimum seasonal occurrence} threshold. The number of seasons of a pattern $P$ is the size of $\mathcal{PS}$, and is computed as $\textit{seasons}(P) = \mid \mathcal{PS} \mid$. 

\textbf{Mining Frequent Seasonal Temporal Patterns} \textbf{(FreqSTP).} Let $\mathcal{D}_{\text{SEQ}}$ be the temporal sequence database, and $\textit{maxPeriod}$, $\textit{minDensity}$, $\textit{distInterval}$, and $\textit{minSeason}$ be the maximum period, minimum density, distance interval, and minimum seasonal occurrence thresholds, respectively. The FreqSTP problem aims to find all frequent seasonal temporal patterns $P$ in $\mathcal{D}_{\text{SEQ}}$ that satisfy the $\textit{maxPeriod}$, $\textit{minDensity}$, $\textit{distInterval}$, and $\textit{minSeason}$ constraints. 
%\vspace{-0.15in}
\section{Distributed Seasonal Temporal Pattern Mining} \label{sec:FTPMfTSMining}

In this section, we provide a distributed algorithm to mine frequent seasonal temporal patterns (\textbf{DSTPM}). The process consists of two steps:  \textit{Seasonal Single Event Mining} and \textit{Seasonal k-Event Pattern Mining} (\text{k} $\geq$ $2$). 
Before introducing the DSTPM algorithm in detail, we first present \textit{candidate seasonal pattern}, a concept designed to support Apriori-like pruning.

\vspace{-0.05in}
%\vspace{-0.05in} 
\subsection{Candidate Seasonal Pattern}\label{sec:Candidate}\vspace{-0.02in}
%The SFTPMfTS problem is to find all temporal patterns that are frequent (i.e., satisfy the support and confidence thresholds) and seasonal (i.e., satisfy the seasonal recurrence threshold). The two measures \textit{support} and \textit{confidence} fulfill the anti-monotonic property \cite{hdfs}, that helps reduce the search space of frequent temporal patterns mining. However, the \textit{seasonal recurrence} measure does not satisfy this property. 
	
Pattern mining methods often use the \textit{anti-monotonicity} property of the \textit{support} measure to reduce the search space \cite{hdfs}. This property ensures that an infrequent event $E_i$ cannot form a frequent 2-event pattern $P$, since support($E_i$) $\ge$ support($P$). Hence, if $E_i$ is infrequent, we can safely remove $E_i$ and any of its combinations from the search space, and still guarantee the algorithm completeness.
However, seasonal temporal patterns constrained by the $\textit{maxPeriod}$, $\textit{minDensity}$, $\textit{distInterval}$ and $\textit{minSeason}$ thresholds do \textit{not} uphold this property, as illustrated below. 

Consider an event $E$ = M:1 and a 2-event pattern $P$ = \text{M:1} $\succcurlyeq$ \text{I:1} in Table \ref{tbl:SequenceDatabase}. Let \textit{maxPeriod} = 2, \textit{minDensity} = 3, \textit{distInterval} = [4, 10], and \textit{minSeason} = 2. From the constraints, we can identify Ithe seasons of $E$ and $P$ as: $\mathcal{PS}^{E}$ = $\{\text{NearSUP}_1^{E}\}$ = $\{G_1,G_2,G_3,G_4,G_5,G_6,G_8,G_9,G_{10},G_{11},G_{13}\}$, and $\mathcal{PS}^{P}$ = $\{\{\text{NearSUP}_1^{P}\}$ = $\{G_1,G_3,G_4,G_5,G_6\}$, $\{\text{NearSUP}_2^{P}\}$ = $\{G_{10},G_{11},G_{13}\}\}$.  Hence, we have: $\mid$$\mathcal{PS}^{E}$$\mid$$=$$1$ and $\mid$$\mathcal{PS}^{P}$$\mid$$=$$2$. Due to the $\textit{minSeason}$ constraint, $E$ is not a frequent seasonal event, whereas $P$ is. This shows that seasonal temporal patterns do not adhere to the anti-monotonic property. 

To improve DSTPM performance, we use the \textit{maximum seasonal occurrence} measure \cite{hoseasonal2023mining}, called $\textit{maxSeason}$, that upholds the anti-monotonicity property.
	
%\textbf{Maximum seasonal occurrence of a single event $E$:} is the ratio between the number of granules in the support set SUP$^E$ of $E$ in $\mathcal{D_{\text{SEQ}}}$, and the $\textit{minDensity}$ threshold:
%\vspace{-0.1in}
%	\begin{equation}
%		\small
%		\textit{maxSeason}({E}) = \frac { |SUP^E|  }{\textit{minDensity}} 
%		\label{eq:maxSeasonEvent}
%	\end{equation}
%Eq. \eqref{eq:maxSeasonEvent} divides the number of granules containing $E$ by the minimum density of a season. Thus, it computes the maximum seasons a single event $E$ can have, given $\mathcal{D_{\text{SEQ}}}$. Similarly, the maximum seasonal occurrence of a group of events $(E_i, ..., E_k)$, denoted as $\textit{maxSeason}(E_i, ..., E_k)$, is the ratio between the number of granules in the support set of $(E_i, ..., E_k)$ and the $\textit{minDensity}$.
	
\hspace{-0.15in}\textbf{Maximum seasonal occurrence of a temporal pattern $P$:} is the ratio between the number of granules in the support set SUP$^P$ of $P$, and the $\textit{minDensity}$ threshold:
\vspace{-0.1in}
	\begin{equation}
		\small
		\textit{maxSeason}({P}) = \frac { |SUP^P| }{\textit{minDensity}} 
		\label{eq:maxSeasonPattern}
	\end{equation}

\begin{lem} \label{lempattern1} \vspace{-0.05in}
		Let $P$ and $P^{'}$ be two temporal patterns such that $P^{'} \subseteq P$. Then $\textit{maxSeason}(P^{'}) \geq \textit{maxSeason}(P)$. 
\end{lem} 

\begin{lem}\label{lemeventpattern}\vspace{-0.05in}
	Let $P$ be a k-event temporal pattern formed by a k-event group $(E_1,...,E_k)$. Then, $\textit{maxSeason}(P) \leq \textit{maxSeason}{(E_1,...,E_k)}$. 
\end{lem} 

%\begin{proof}
%	Detailed proofs of all lemmas, theorems, and complexities in this article can be found in the technical report \cite{ho2022seasonal}.
%\end{proof}

%From Lemma \ref{lempattern1}, the $\textit{maxSeason}$ of a group of events $G_E$ (or pattern $P$) is always at most the $maxSeason$ of its subset $G_E^{'}$ (or sub-pattern $P^{'}$). \textcolor{blue}{From Lemma \ref{lemeventpattern}, the $maxSeason$ of a pattern $P$ is at most the $maxSeason$ of its events.} Thus, $maxSeason$ upholds the anti-monotonicity property, and can be used to reduce the search space of STPM. Below, we define the \textit{candidate pattern} concept that uses $maxSeason$ as a gatekeeper to identify frequent/ infrequent seasonal patterns.
The $\textit{maxSeason}$ upholds the anti-monotonicity property \cite{hoseasonal2023mining}, and can be used to reduce the DSTPM search space. Below, we define the \textit{candidate pattern} concept that uses $\textit{maxSeason}$ as a gatekeeper to identify frequent/ infrequent seasonal patterns.
	
\hspace{-0.15in}\textbf{Candidate seasonal pattern:} A temporal pattern $P$ is a \textit{candidate seasonal pattern} if $\textit{maxSeason}(P) \geq \textit{minSeason}$.

Similarly, a group of k events $G_E= (E_1,..., E_k)$ ($k \ge 1$) is a \textit{candidate seasonal k-event group} if $\textit{maxSeason}(G_E) \geq \textit{minSeason}$. %A candidate seasonal 1-event group is also called the \textit{candidate seasonal single event}.
%Intuitively, a pattern $P$ (or k-event group $G_E$) can potentially be a frequent seasonal pattern (or k-event group) if its $\textit{maxSeason}$ is greater than the predefined minimum seasonal occurrence threshold. In contrast, a pattern $P$ (or $G_E$) cannot be \textcolor{blue}{a frequent seasonal pattern (or k-event group)} if its $\textit{maxSeason}$ is less than $\textit{minSeason}$. Hence, $P$ (or $G_E$) can be safely removed from the search space.
Intuitively, a pattern $P$ (or k-event group $G_E$) is infrequent if its $\textit{maxSeason}$ is less than $\textit{minSeason}$. Hence, $P$ (or $G_E$) can be safely removed from the search space.

Next, we present our DSTPM algorithm and detail the two mining steps. Algorithm \ref{algorithmSFTPM} provides the pseudo-code of DSTPM. %We detail each mining step in the next section.

\vspace{-0.05in} \subsection{Mining Seasonal Single Events} \label{sec:1patternmining}

\setlength{\textfloatsep}{0pt}
\SetNlSty{}{}{:} 	%\vspace{-0.1in}	%set number line in term of 1: , 2:
\begin{algorithm}[!t]
	\algsetup{linenosize=\tiny}
	\SetInd{0.5em}{0.5em}
	\small
	%\scriptsize
	\DontPrintSemicolon
	\caption{\mbox{Distributed Frequent Seasonal Temporal Pattern Mining}}
	\label{algorithmSFTPM}
	
	\KwInput{Temporal sequence database $\mathcal{D_{\text{SEQ}}}$, the thresholds: $\textit{maxPeriod}$, $\textit{minDensity}$, $\textit{distInterval}$, $\textit{minSeason}$}
	\KwOutput{The set of frequent seasonal temporal patterns $\mathcal{P}$}
	
	\nonl //Mining frequent seasonal single events \;
	DEvent $\leftarrow$ RetrieveDistributedEvent($\mathcal{D}_{\text{SEQ}}$);\;
	Candidate1Event $\leftarrow$ FilterCandidate\_maxSeason(DEvent);\;
	FSEvent $\leftarrow$ FindFrequentSeasonalEvents($\textit{maxPeriod}$, $\textit{minDensity}$, $\textit{distInterval}$, $\textit{minSeason}$);\;
	
	\nonl //Mining frequent seasonal k-event patterns ($k \geq 2$)\;
	kEventGroups $\leftarrow$ Cartesian($F_1$, $F_{k-1}$);\;  
	CandidatekEvent $\leftarrow$ FilterCandidateEvent\_maxSeason(kEventGroups);\;
	CandidatekPattern $\leftarrow$ Retrieve\_FilterCandidatePattern(CandidatekEvent);\;
	FSPattern $\leftarrow$ FindFrequentSeasonalPatterns($\textit{maxPeriod}$, $\textit{minDensity}$, $\textit{distInterval}$, $\textit{minSeason}$);\;
\end{algorithm}

	\iffalse
	%Compute measure for seasonal recurrence
	\SetNlSty{}{}{:} 	%\vspace{-0.1in}	%set number line in term of 1: , 2:
	\begin{algorithm}
		\algsetup{linenosize=\tiny}
		\SetInd{0.5em}{0.5em}
		\small
		\DontPrintSemicolon
		\caption{Calculate seasonal occurrence $SR$ (ComputeSR)}
		\label{algorithmSR}
		
		\KwInput{A support set $SUP$, a maximum period threshold $maxPer$, a minimum density $minDen$, a minimum seasonal occurrence $minSea$, a distance interval $[dist_{min}, dist_{max}]$}
		\KwOutput{The value of seasonal occurrence $SR$}
		
		$PS$ = [[0]]; $NearSUP = None$; \;
		\ForEach{\textit{granule} $h \in SUP$}{
			\nonl // Looking for the first element of a NearSUP \;
			\If{NearSUP is None}{ 
				$distance = p(h) - p(PS[-1][-1])$ ; // PS[-1][-1] is the last granule of the lastest NearSUP in PS  \; 
				\If{$distance$ is in $[dist_{\min}, dist_{\max}]$}{
					$NearSUP = [h]$ ; \;
				}
				continue;\;
			}
			\nonl // Checking period between two granules \;
			$period = h - NearSUP[-1]$; // NearSUP[-1] is the last granule in NearSUP \;
			\If{$period \leq maxPer$}{
				Insert $h$ into $NearSUP$ ;\; 
			}
			\Else{
				$density = |NearSUP|$; \;
				\If{$density \geq minDen$}{
					Insert $NearSUP$ into $PS$ ;\; 
				}
				$distance = p(h) - p(PS[-1][-1])$ ; \;
				\If{$distance$ is in $[dist_{min}, dist_{max}]$}{
					$NearSUP = [h]$ ;\;
				}
				\Else{
					$NearSUP = None$ ;\;
				}
			}
		}
		$SR = |PS|$ ; \;		
	\end{algorithm}
	\fi

The first step in DSTPM is to mine frequent seasonal single events (Alg. \ref{algorithmSFTPM}, lines 1-3) that satisfy the constraints of $\textit{maxPeriod}$, $\textit{minDensity}$, $\textit{distInterval}$ and $\textit{minSeason}$. To do that, we first look for the candidate single events defined in Section \ref{sec:Candidate}, and then use only the found candidates to mine frequent seasonal events.

\begin{figure*}[!b]
	%\vspace{-0.6in}
	\setlength{\tabcolsep}{0pt}
	\begin{tabularx}{\linewidth}{ll}
		\begin{minipage}{.35\linewidth}
			\begin{minipage}{\linewidth}
				\captionsetup{justification=centering, font=small}
				\includegraphics[width=1\textwidth,height=3.1cm]{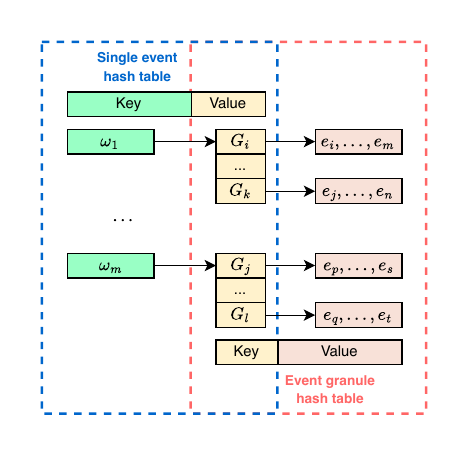}
				\vspace{-0.2in}
				\caption{The $HLH_1$ structure}
				\label{fig:hlh1}
			\end{minipage}
		\end{minipage}&
		\begin{minipage}{.66\linewidth}
			\begin{minipage}{\linewidth}
				\centering
				\captionsetup{justification=centering, font=small}
				\includegraphics[width=1\textwidth,height=3.1cm]{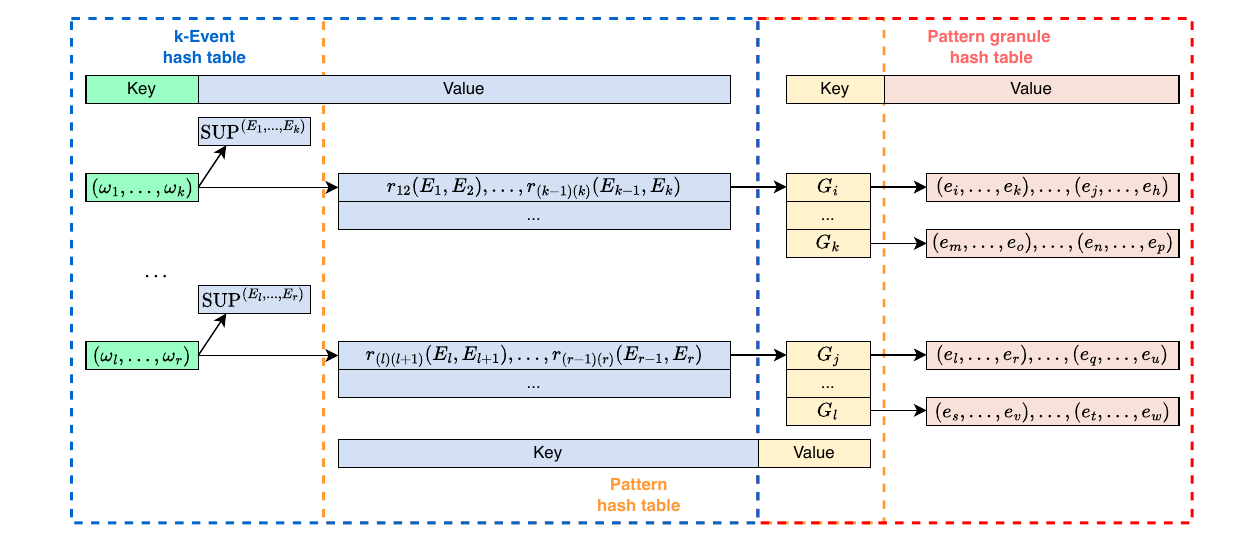}
				\vspace{-0.2in}
				\caption{The $HLH_k (k \geq 2)$ structure}
				\label{fig:hlhk}
			\end{minipage}
		\end{minipage}
	\end{tabularx}
	\vspace{-0.2in}
\end{figure*}

We use a \textit{distributed hierarchical lookup hash structure} $DHLH_1$ to store the candidate seasonal single events. This data structure enables fast search when mining seasonal k-events patterns ($k \geq 2$). Note that we maintain the candidate events in $DHLH_1$ instead of the frequent seasonal events, as the $\textit{maxSeason}$ of candidate events upholds the anti-monotonicity property, and can thus be used for pruning. We illustrate $DHLH_1$ in Fig. \ref{fig:hlh1}, and describe the data structure below.

\textbf{Distributed hierarchical lookup hash structure $DHLH_1$:} The $DHLH_1$ is a hierarchical data structure that consists of two hash tables: the \textit{single event hash table} $EH$,  
and the \textit{event granule hash table} $GH$.
Each hash table has a list of $<$key, value$>$ pairs. 
In $EH$, the key is the event symbol $\omega \in \Sigma_X$ representing the candidate $E_i$, and the value is the list of granules $<G_i,...,G_k>$ in $SUP^{E_i}$. In $GH$, the key is the list of granules shared in the value field of $EH$, while the value stores event instances of $E_i$ that appear at the corresponding granule in $\mathcal{D}_{\text{SEQ}}$.
The $HLH_1$ structure enables fast retrieval of event granules and instances when mining candidate seasonal k-event patterns in the next step of DSTPM. 

\begin{figure}[!t]
	\centering
	\captionsetup{justification=centering, font=small}
	\begin{minipage}{\linewidth}
		\includegraphics[width=\textwidth]{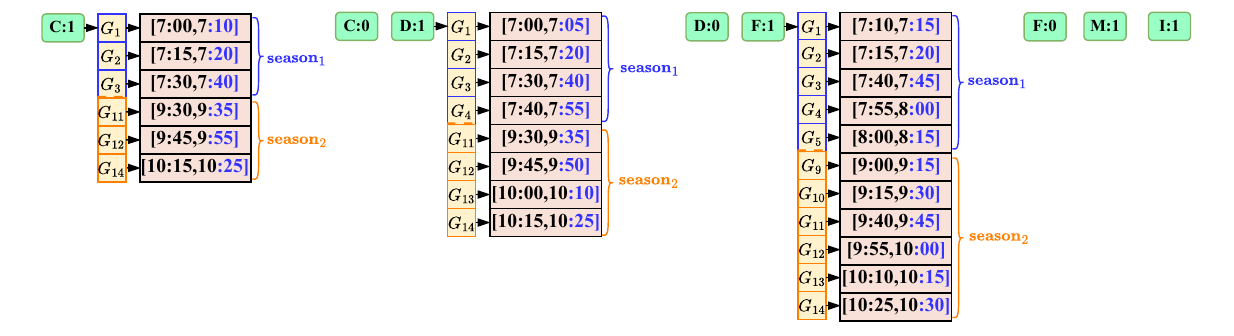}
		\vspace{-0.3in}
		\caption{A hierarchical lookup hash tables $DHLH_1$}
		\label{fig:exhlh1}
	\end{minipage}
\end{figure}

The process of mining frequent seasonal events is as follows. First, $\mathcal{D}_{\text{SEQ}}$ is loaded into a Spark RDD, and a \textit{map()} function is used to split the columns of $\mathcal{D}_{\text{SEQ}}$, transforming each row into a (key, value) format (Alg. \ref{algorithmSFTPM}, line 1). Here, the key is the event symbol $\omega \in \Sigma_X$ representing the candidate $E_i$, and the value is the list of granules and its event instances. Next, a \textit{reduceByKey()} function is used to aggregate the values that share the same key, i.e., same event symbol. Then, a \textit{map()} function is used to extract the support set $SUP^{E_i}$ for each event $E_i$, from which we compute the maximum seasonal occurrence $\textit{maxSeason}(E_i)$. Next, a \textit{filter()} function is used to select candidate seasonal single events with $\textit{maxSeason}(E_i) \geq \textit{minSeason}$ (Alg. \ref{algorithmSFTPM}, line 2). To mine frequent seasonal events, for each candidate event $E_i$, the support set $SUP^{E_i}$ to calculate the period $pr_{ij}$ between every two consecutive granules in $SUP^{E_i}$, and determine the near support sets $\text{NearSUP}^{E_i}$ that satisfy $\textit{maxPeriod}$ and $\textit{minDensity}$.  Next, the set of seasons $\mathcal{PS}^{E_i}$ is identified by selecting the near support sets that adhere to the $\textit{distInterval}$ constraint. Then, the frequent seasonal events are determined using the \textit{filter()} function by  selecting only those that have $\textit{seasons}(E_i) \geq \textit{minSeason}$ (Alg. \ref{algorithmSFTPM}, line 3). Finally, the frequent seasonal events are mapped to the $DHLH_1$.

We provide an example of $DHLH_1$ in Fig. \ref{fig:exhlh1} using data in Table \ref{tbl:SequenceDatabase} with \textit{maxPeriod} = 2, \textit{minDensity} = 3, \textit{distInterval} = [4, 10], and \textit{minSeason} = 2. Here, out of $10$ events in $\mathcal{D}_{\text{SEQ}}$, we have eight candidate seasonal single events stored in $DHLH_1$: C:1, C:0, D:1, D:0, F:1, F:0, M:1, and I:1. Due to space limitations, we only provide the detailed internal structure of three candidate events. Among the eight candidates, the event M:1 does not satisfy the $\textit{minSeason}$ threshold since $\textit{season}$(M:1) = 1, and thus, is not a frequent seasonal event. However, M:1 is still present in $DHLH_1$ as M:1 might create frequent seasonal k-event patterns. In contrast, I:0 and M:0 are not the candidate seasonal events because they do not satisfy the $\textit{maxSeason}$ constraint, and are omitted from $DHLH_1$.
\vspace{-0.1in}
\subsection{Mining Seasonal k-event Patterns} \label{sec:kpatternmining}
Below, we introduce the distributed hierarchical lookup hash structure $HLH_k$ used to store candidates seasonal k-event patterns. 

\textbf{The distributed hierarchical lookup hash structure $DHLH_k$:} We use the \textit{hierarchical lookup hash structure} $DHLH_k$ $(k \geq 2)$ to maintain candidate seasonal k-event groups and patterns, as illustrated in Fig. \ref{fig:hlhk}. 
The $DHLH_k$ contains three hash tables: the \textit{k-event hash table} $EH_k$, the \textit{pattern hash table} $PH_k$, and the \textit{pattern granule hash table} $GH_k$. 
%\sout{The $EH_k$ table maintains the information of candidate k-event groups, and their corresponding k-event patterns. Thus,} 
For each {\em $<$key, value$>$} pair of $EH_k$, {\em key} is the list of symbols $(\omega_1 ..., \omega_k)$ representing the candidate k-event group $(E_1,...,E_k)$, and {\em value} is an \textit{object} which consists of two components: (1) the support set $SUP^{(E_1,...,E_k)}$, and (2) a list of candidate seasonal k-event temporal patterns. %$P = \{(r_{12},E_1,E_2), ..., (r_{(k-1)(k)},E_{k-1},E_{k})\}$, created from the k-event group $(E_1,...,E_k)$.  
In $PH_k$, {\em key} is the candidate pattern $P$ which takes the {\em value} component of $EH_k$, while {\em value} is the list of granules that contain $P$. %\sout{Finally, the $GH_k$ table maintains the event instances involved in pattern $P$.} 
In $GH_k$, {\em key} is the list of granules containing $P$ which takes the {\em value} component of $PH_k$, while {\em value} is the list of event instances from which the temporal relations in $P$ are formed. 
The $DHLH_k$ helps speed up the candidate seasonal k-event group mining through the use of the support set in $EH_k$, and enables fast search for temporal relations between $k$ events using the information in $PH_k$ and $GH_k$.

\textbf{4.1 Mining candidate seasonal k-event groups.} We first find candidate seasonal k-event groups (Alg. \ref{algorithmSFTPM}, lines 4-5).
 
Let $F_{k-1}$ be the set of candidate seasonal (k-1)-event groups found in $DHLH_{k-1}$, and $F_1$ be the set of candidate seasonal single events in $DHLH_1$. We first generate all possible k-event groups by computing the Cartesian product $F_{k-1} \times F_1$. Next, for each k-event group $(E_1, ..., E_k)$, we use a \textit{map()} function to select  the support set $SUP^{(E_1, ..., E_k)}$ from $SUP^{(E_1, ..., E_{k-1})}$ in $EH_{k-1}$ and $SUP^{E_k}$ in $EH$, and then use a \textit{filter()} function to select the support set $SUP^{(E_1, ..., E_k)}$ and $\textit{maxSeason}(E_1, ..., E_k)$, and keep $(E_1, ..., E_k)$ if it is a candidate, and it is mapped to the $EH_k$ of $HLH_k$.

\textbf{4.2 Mining frequent seasonal k-event patterns.} We use the candidate k-event groups to mine frequent seasonal k-event patterns (Alg. \ref{algorithmSFTPM}, lines 6-7).  
%We first discuss the case of 2-event patterns, and then generalize to k-event patterns.

%\textit{4.2.1 Mining frequent seasonal 2-event patterns:} 
%For each candidate 2-event group $(E_i,E_j)$, we use the support set $SUP^{(E_i,E_j)}$ to retrieve the temporal sequences $\mathcal{S}$ that contain $(E_i,E_j)$. Next, for each sequence $S \in \mathcal{S}$, we extract their event instances $(e_i,e_j)$, and verify the relation between them. We then compute the $\textit{maxSeason}$ of the 2-event pattern $P$ and determine if $P$ is a candidate pattern, i.e., $\textit{maxSeason}(P) \geq \textit{minSeason}$. Finally, the candidate seasonal 2-event patterns are stored in $PH_2$, while their event instances are stored in $GH_2$.

%Based on the set of candidate seasonal 2-event patterns $P$, we determine whether $P$ is a frequent seasonal 2-event pattern by checking the constraints of $\textit{maxPeriod}$, $\textit{minDensity}$, $\textit{distInterval}$ and $\textit{minSeason}$ as in the case of single events, using the support set $SUP^P$ retrieved from the value of $PH_2$. 

%\textit{4.2.2 Mining frequent seasonal k-event patterns:} 
Let $N_{k-1}=(E_1,...,E_{k-1})$ be a candidate (k-1)-event group in $DHLH_{k-1}$, $N_1=(E_k)$ be a candidate single event in $DHLH_1$, and $N_k=N_{k-1} \cup N_1 = (E_1,...,E_k)$ be a candidate k-event in $DHLH_k$. 
To find k-event patterns for $N_k$, we first retrieve the set of candidate (k-1)-event patterns $\mathcal{P}_{k-1}$. Each $P_{k-1} \in \mathcal{P}_{k-1}$ is a list of $\frac{1}{2}(k-1)(k-2)$ triples: $\{(r_{12}$, $E_{1}$, $E_{2})$,...,$(r_{(k-2)(k-1)}$, $E_{k-2}$, $E_{k-1})\}$. We iteratively verify the possibility of $P_{k-1}$ forming a k-event pattern $P_k$ with $E_k$ as follows.  
We first start with the triple $(r_{(k-1)k}$, $E_{k-1}$, $E_{k})$. If $(r_{(k-1)k}$, $E_{k-1}$, $E_{k})$ does not exist in $DHLH_2$, then $P_k$ is not a candidate k-event pattern, and the verification stops immediately. Otherwise, we continue the similar verification on the triple $(r_{(k-2)k}$, $E_{k-2}$, $E_k)$, until it reaches $(r_{1k}$, $E_{1}$, $E_{k})$. Next, we use a \textit{filter()} function to select candidate k-event patterns. The candidate k-event patterns are mapped in $PH_k$ and $GH_k$. 
Based on the candidate seasonal k-event patterns $P$, we determine whether a  $P$ is a frequent seasonal k-event pattern by using a \textit{filter()} function to check the constraints of $\textit{maxPeriod}$, $\textit{minDensity}$, $\textit{distInterval}$ and $\textit{minSeason}$ as in the case of single events.

\begin{figure}[!t]
	\centering
	\captionsetup{justification=centering, font=small}
	\begin{minipage}{\linewidth}
		\includegraphics[width=\textwidth]{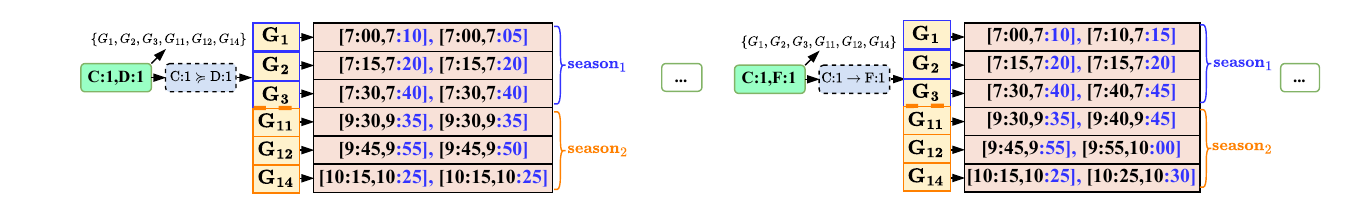}
		\vspace{-0.3in}
		\caption{A hierarchical lookup hash tables $DHLH_2$}
		\label{fig:exhlh2}
	\end{minipage}
\end{figure}

Fig. \ref{fig:exhlh2} is an example of frequent seasonal 2-event patterns at $DHLH_2$ using data in Table \ref{tbl:SequenceDatabase}. The patterns $P_1$ = \text{C:1} $\succcurlyeq$ \text{D:1} and $P_2$ = \text{C:1} $\rightarrow$ \text{F:1} are frequent seasonal 2-event patterns.

\vspace{-0.2in}\section{Experimental Evaluation}\label{sec:experiment}
\subsection{Experimental Setup} \label{sec:ExperimentalSetup} \vspace{-0.02in}
\textbf{Datasets:}
We use three real-world datasets from three application domains: renewable energy \cite{renewableEnergy}, smart city \cite{smartcity}, and health \cite{diseasedata}. The  \textit{renewable energy} (RE) measures the energy consumption of electrical appliances in residential households from Spain. The \textit{smart city} (SC) is traffic data from New York City. The \textit{health} is data of the influenza (INF) disease from Kawasaki, Japan. 
 
%\hspace{-0.2in}\textbf{Baseline method:}  Our method is referred to as DSTPM.  
%Since there are no existing methods performing distributed seasonal temporal pattern mining, we compare performance (both runtime and memory usage) between DSTPM and the adaped PS-growth (APS). PS-growth \cite{kiran2019finding} mines recurring itemset, we since adapt it to find seasonal temporal patterns. Specifically, the adaptation is done through 2-phase process: (1) PS-growth is applied to find frequent recurring events, and (2), mine temporal patterns from extracted events. The adapted PS-growth is referred to as APS. 
  
\hspace{-0.2in}\textbf{Infrastructure:}
The experiments are conducted at our research laboratory with a cluster of 30 high-end nodes, each with 64 cores, 256GB of RAM, and 10TB of storage.
  
\hspace{-0.2in}\textbf{Parameters:} Table \ref{tbl:params} lists the parameters and their values used in our experiments, where \textit{maxPeriod} and \textit{minDensity} are expressed as the percentage of $\mathcal{D}_{\text{SEQ}}$. 

\begin{table*}[!t]	
	%\vspace{-0.1in}
	\centering
	\begin{minipage}{.69\linewidth} %.25
		\captionsetup{justification=centering, font=small}
		\caption{\small Characteristics of the Datasets}
		\small
		\resizebox{\columnwidth}{1.03cm}{
			\begin{tabular}{ |c|c|c|c|c|} 
				\hline {\bfseries Datasets} & {\bfseries \#seq.} & {\bfseries \#time series} & {\bfseries \#events} & {\bfseries \#ins./seq.} \\ 
				\hline
				RE (real)  &  1,460 &  21 &  102 &   93\\
				\hline 
				\centering SC (real) &  1,249 &  14 &  56 &  55\\
				\hline 
				\centering INF (real) &  608 &  25 &  124 &  48\\
				\hline 
				\centering HFM (real) &  730 &  24 &  115 &  40\\
				\hline
			\end{tabular} 
		}			
		\label{tbl:datasetCharacteristic}
	\end{minipage}
	%\hspace{0.1in}
	\begin{minipage}{.3\linewidth} %.72
		\captionsetup{justification=centering, font=small}
		\caption{\small Parameters}
		\small
		\resizebox{\columnwidth}{1.03cm}{
			\begin{tabular}{ |l|c| }
				\hline &\\[0.01em]  \thead{Params} & {\bfseries Values (User-defined)}  \\ &\\[0.01em]
				%\hline {\bfseries Params} & {\bfseries Values (User-defined)} \\ 
				\hline 
				$\textit{maxPeriod}$ & 0.2\%, 0.4\%, 0.6\%, 0.8\%, 1.0\% \\
				\hline
				$\textit{minDensity}$ & 0.5\%, 0.75\%, 1.0\%, 1.25\%, 1.5\% \\
				\hline
				$\textit{minSeason}$ & 4, 8, 12, 16, 20 \\
				\hline
				$\textit{distInterval}$ &  
				[90, 270] (RE, SC), [30, 90] (INF, HFM) \\
				\hline
			\end{tabular}
		}			
		\label{tbl:params}
	\end{minipage}
	\vspace{-0.2in}
\end{table*}

\subsection{Qualitative Evaluation}%\vspace{-0.02in}
Table \ref{tbl:interestingPatterns} lists some seasonal patterns found in the datasets.
Patterns P1-P2 are extracted from RE, showing that high renewable energy generation and high electricity demand occur seasonally and often at specific \textit{season} throughout the year. 
Patterns P3-P4 are extracted from INF, showing the detection of seasonal diseases. 
Finally, how weather affects traffic is shown in patterns P5-P6 extracted from SC.
	
\begin{table}[!t]	
	%\vspace{-0.1in}
		\captionsetup{justification=centering, font=small}
		\caption{\small Summary of Interesting Seasonal Patterns}
		\centering
		\resizebox{\textwidth}{1.25cm}{ %2.3cm
			\begin{tabular}{ |l|c|c|c|c|} %p{11cm}
				\hline &&&&\\[-1em]  {\bfseries \;\;\;\;\;\;\;\;\;\;\;\;\;\;\;\;\;\;\;\;\;\;\;\;\;\;\;\;\;\;\;\;\;\;\;\;\;\;\;\;\;\;\;\;\;\;\;\;\;\;\;\;\;\;\;\;\;\;\;\;\;\;\;\;\;\;\;\;\;\;\;\;\;\; \small Patterns} & {\bfseries \small minDensity (\%)} & {\bfseries \small  maxPeriod (\%)} & {\bfseries \small   \# minSeason} & {\bfseries \small  Seasonal occurrence} \\
				\hline &&&&\\[-1em]
				\normalsize (P1) Low Temperature $\succcurlyeq$ High Energy Consumption & \normalsize 0.5 & \normalsize 0.4 & \normalsize 12 & \normalsize December, January, February \\			
				\hline   &&&&\\[-1em]
				\normalsize (P2) Very Few Clouds $\succcurlyeq$ High Temperature $\between$ High Solar Power Generation & \normalsize 0.75 & \normalsize  0.6 & \normalsize  8 & \normalsize  July, August \\
				\specialrule{1.5pt}{1pt}{1pt}
				\normalsize  (P3) High Humidity $\between$ Very Low Temperature $\rightarrow$ Very High Influenza Cases  & \normalsize  0.5 & \normalsize  0.4 & \normalsize  12 & \normalsize  January, February \\
				\hline   &&&&\\[-1em]
				\normalsize  (P4) Strong Wind $\succcurlyeq$ Heavy Rain $\succcurlyeq$ High Influenza Cases & \normalsize  0.5 & \normalsize  0.4 & \normalsize 12 & \normalsize  January, February \\
				\specialrule{1.5pt}{1pt}{1pt}  
				\normalsize  (P5) High Temperature $\succcurlyeq$ Strong Wind $\rightarrow$ High Congestion  & \normalsize 0.5 & \normalsize 0.6 & \normalsize 8 & \normalsize  July, August \\
				\hline  &&&&\\[-1em]
				\normalsize  (P6) Heavy Rain $\succcurlyeq$ Unclear Visibility  $\succcurlyeq$ High Lane-Blocked  & \normalsize 0.4 & \normalsize 0.8 & \normalsize 8 & \normalsize  July, August \\
				\hline     
			\end{tabular} 
		}
		\label{tbl:interestingPatterns}
\end{table}

%\subsection{Quantitative Evaluation}%\vspace{-0.02in}

\subsection{Baseline comparison on real-world datasets}\label{sec:baselinecomparison}

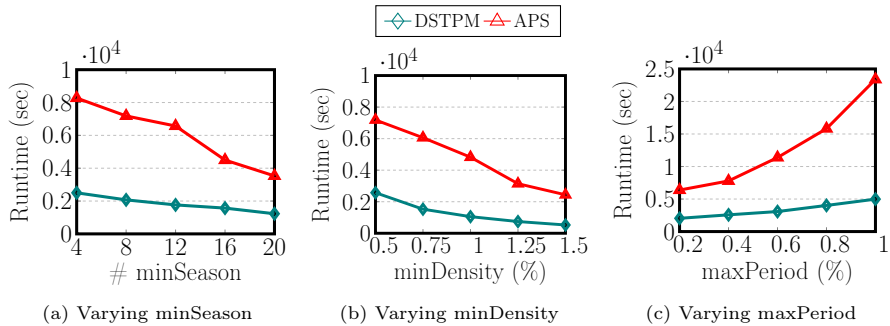
\begin{figure*}[!t]
	\vspace{-0.1in}
	\hspace{1.95in}
	\ref{legendcomparison}
	\clearpage
	\vspace{-0.15in}
	\begin{minipage}[t]{1\columnwidth} 
		\centering
		\begin{subfigure}{0.32\columnwidth}
			\centering
			\resizebox{\linewidth}{!}{
				\begin{tikzpicture}[scale=0.6]
					\begin{axis}[
						compat=newest,
						xlabel={\# minSeason},
						ylabel={Runtime (sec)}, 
						label style={font=\Huge},
						ticklabel style = {font=\Huge},
						xmin=4, xmax=20,
						ymin=0, ymax=10000,
						xtick={4,8,12,16,20},
						ytick ={0,2000,4000,6000,8000,10000},
						legend columns=-1,
						legend entries = {DSTPM, APS},
						legend image post style={scale=0.6},  %%% scale legend marker size
						legend style={nodes={scale=0.5,  transform shape}, font=\Large},
						legend to name={legendcomparison},
						%ymode=log,
						%log basis y={10},
						%log ticks with fixed point,
%						y tick label style={
%							/pgf/number format/.cd,
%							fixed,
%							precision=2,
%							/tikz/.cd
%						},
						ymajorgrids=true,
						grid style=dashed,
						line width=3pt
						]
						\addplot[
						color=teal,
						mark=diamond,
						mark size=5pt,
						]
						coordinates {
							(4,2501)(8,2072)(12,1765)(16,1570)(20,1227)
						};
						\addplot[
						color=red,
						mark=triangle,
						mark size=5pt,
						]
						coordinates {
							(4,8273)(8,7183)(12,6570)(16,4496)(20,3530)
						};						
					\end{axis}
				\end{tikzpicture}
			}
		\captionsetup{justification=centering, font=scriptsize}
			\caption{\scriptsize Varying minSeason}
		\end{subfigure}
		\begin{subfigure}{0.32\columnwidth}
			\centering
			\resizebox{\linewidth}{!}{
				\begin{tikzpicture}[scale=0.6]
					\begin{axis}[
						compat=newest,
						xlabel={minDensity (\%)},
						ylabel={Runtime (sec)}, 
						label style={font=\Huge},
						ticklabel style = {font=\Huge},
						xticklabel style = {xshift=3mm}, %shift x tick label
						xmin=0.5, xmax=1.5,
						ymin=0, ymax=10000,
						xtick={0.5,0.75,1,1.25,1.5},
						ytick ={0,2000,4000,6000,8000,10000},
						legend columns=-1,
						legend entries =  {DSTPM, APS},
						legend image post style={scale=0.6},  %%% scale legend marker size
						legend style={nodes={scale=0.5,  transform shape}, font=\Large},
						legend to name={legendcomparison},
						%ymode=log,
						%log basis y={10},
						ymajorgrids=true,
						grid style=dashed,
						line width=3pt
						]				
						\addplot[
						color=teal,
						mark=diamond,
						mark size=5pt,
						]
						coordinates {
							(0.5,2572)(0.75,1526)(1,1053)(1.25,743)(1.5,526)
						};						
						\addplot[
						color=red,
						mark=triangle,
						mark size=5pt,
						]
						coordinates {
							(0.5,7183)(0.75,6059)(1,4824)(1.25,3142)(1.5,2439)
						};
					\end{axis}
				\end{tikzpicture}
			}
		\captionsetup{justification=centering, font=scriptsize}
			\caption{\scriptsize Varying minDensity}
		\end{subfigure}
		\begin{subfigure}{0.32\columnwidth}
			\centering
			\resizebox{\linewidth}{!}{
				\begin{tikzpicture}[scale=0.6]
					\begin{axis}[
						compat=newest,
						xlabel={maxPeriod (\%)},
						ylabel={Runtime (sec)}, 
						label style={font=\Huge},
						ticklabel style = {font=\Huge},
						xticklabel style = {xshift=3mm}, %shift x tick label
						xmin=0.2, xmax=1,
						ymin=0, ymax=25000,
						xtick={0.2,0.4,0.6,0.8,1},
						ytick={0,5000,10000,15000,20000,25000},
						legend columns=-1,
						legend entries =  {DSTPM, APS},
						legend image post style={scale=0.6},  %%% scale legend marker size
						legend style={nodes={scale=0.5,   transform shape}, font=\Large},
						legend to name={legendcomparison},
						%ymode=log,
						%log basis y={10},
						ymajorgrids=true,
						grid style=dashed,
						line width=3pt
						]			
						\addplot[
						color=teal,
						mark=diamond,
						mark size=5pt,
						]
						coordinates {
							(0.2,2035)(0.4,2572)(0.6,3077)(0.8,4015)(1,4984)
						};
						\addplot[
						color=red,
						mark=triangle,
						mark size=5pt,
						]
						coordinates {
							(0.2,6381)(0.4,7783)(0.6,11382)(0.8,15819)(1,23392)
						};
					\end{axis}
				\end{tikzpicture}
			}
		\captionsetup{justification=centering, font=scriptsize}
			\caption{\scriptsize Varying maxPeriod}
		\end{subfigure}
		%\vspace{-0.1in}
		%\ref{legendcomparison}
		\vspace{-0.08in}
		\captionsetup{justification=centering, font=small}
		\caption{Runtime Comparison on RE (real-world)}
		\label{fig:runtimebaselineRE}
	\end{minipage}
\end{figure*}  
	
\begin{figure*}[!t]
	\begin{minipage}[t]{1\columnwidth} 
		\centering
		\begin{subfigure}{0.32\columnwidth}
			\centering
			\resizebox{\linewidth}{!}{
				\begin{tikzpicture}[scale=0.6]
					\begin{axis}[
						compat=newest,
						xlabel={\# minSeason},
						ylabel={Runtime (sec)}, 
						label style={font=\Huge},
						ticklabel style = {font=\Huge},
						xmin=4, xmax=20,
						ymin=0, ymax=10000,
						xtick={4,8,12,16,20},
						ytick ={0,2000,4000,6000,8000,10000},
						legend columns=-1,
						legend entries = {DSTPM, APS},
						legend style={nodes={scale=0.55,  transform shape}, font=\Large},
						legend to name={legendpruning},
						%ymode=log,
						%log basis y={10},
						ymajorgrids=true,
						grid style=dashed,
						line width=3pt
						]				
						\addplot[
						color=teal,
						mark=asterisk,
						mark size=5pt,
						]
						coordinates {
							(4,2018)(8,1938)(12,1630)(16,1338)(20,904)
						};
						\addplot[
						color=red,
						mark=triangle,
						mark size=5pt,
						]
						coordinates {
							(4,7120)(8,6002)(12,5420)(16,4001)(20,2763)
						};						
					\end{axis}
				\end{tikzpicture}
			}
			\caption{\scriptsize Varying minSeason.}
		\end{subfigure}
		\begin{subfigure}{0.32\columnwidth}
			\centering
			\resizebox{\linewidth}{!}{
				\begin{tikzpicture}[scale=0.6]
					\begin{axis}[
						compat=newest,
						xlabel={minDensity (\%)},
						ylabel={Runtime (sec)}, 
						label style={font=\Huge},
						ticklabel style = {font=\Huge},
						xmin=0.5, xmax=1.5,
						ymin=0, ymax=10000,
						xtick={0.5,0.75,1,1.25,1.5},
						ytick ={0,2000,4000,6000,8000,10000},
						legend columns=-1,
						legend entries = {DSTPM, APS},
						legend style={nodes={scale=0.55,  transform shape}, font=\Large},
						legend to name={legendpruning},
						%ymode=log,
						%log basis y={10},
						ymajorgrids=true,
						grid style=dashed,
						line width=3pt
						]				
						\addplot[
						color=teal,
						mark=asterisk,
						mark size=5pt,
						]
						coordinates {
							(0.5,2238)(0.75,1332)(1,725)(1.25,529)(1.5,410)
						};						
						\addplot[
						color=red,
						mark=triangle,
						mark size=5pt,
						]
						coordinates {
							(0.5,7002)(0.75,5501)(1,4084)(1.25,2210)(1.5,1873)
						};
					\end{axis}
				\end{tikzpicture}
			}
			\caption{\scriptsize Varying minDensity.}
		\end{subfigure}
		\begin{subfigure}{0.32\columnwidth}
			\centering
			\resizebox{\linewidth}{!}{
				\begin{tikzpicture}[scale=0.6]
					\begin{axis}[
						compat=newest,
						xlabel={maxPeriod (\%)},
						ylabel={Runtime (sec)}, 
						label style={font=\Huge},
						ticklabel style = {font=\Huge},
						xmin=0.2, xmax=1,
						ymin=0, ymax=20000,
						xtick={0.2,0.4,0.6,0.8,1},
						ytick ={0,5000,10000,15000,20000,25000},
						legend columns=-1,
						legend entries = {DSTPM, APS},
						legend style={nodes={scale=0.55,   transform shape}, font=\Large},
						legend to name={legendpruning},
						%ymode=log,
						%log basis y={10},
						ymajorgrids=true,
						grid style=dashed,
						line width=3pt
						]					
						\addplot[
						color=teal,
						mark=asterisk,
						mark size=5pt,
						]
						coordinates {
							(0.2,1426)(0.4,1838)(0.6,2452)(0.8,2619)(1,2859)
						};
						\addplot[
						color=red,
						mark=triangle,
						mark size=5pt,
						]
						coordinates {
							(0.2,5031)(0.4,6002)(0.6,8931)(0.8,11932)(1,17023)
							%(0.2,ABC)(0.4,ABC)(0.6,ABC)(0.8,ABC)(1,ABC)
						};
					\end{axis}
				\end{tikzpicture}
			}
			\caption{\scriptsize Varying maxPeriod.}
		\end{subfigure}
		\caption{Runtime Comparison on SC (real-world)}
		\label{fig:runtimebaselineSC}
	\end{minipage}%
\end{figure*}
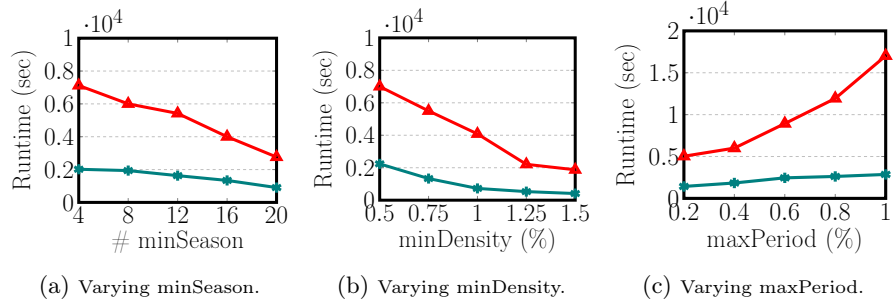  

\begin{figure*}[!t]
	\vspace{-0.1in}
	\begin{minipage}[t]{1\columnwidth} 
		\centering
		\begin{subfigure}{0.32\columnwidth}
			\centering
			\resizebox{\linewidth}{!}{
				\begin{tikzpicture}[scale=0.6]
					\begin{axis}[
						compat=newest,
						xlabel={\# minSeason},
						ylabel={Memory Usage (MB)}, 
						label style={font=\Huge},
						ticklabel style = {font=\Huge},
						xmin=4, xmax=20,
						ymin=0, ymax=15000,
						xtick={4,8,12,16,20},
						ytick ={0, 5000,10000,15000},
						legend columns=-1,
						legend entries = {DSTPM, APS},
						%legend image post style={scale=0.6},  %%% scale legend marker size
						legend style={nodes={scale=0.5,  transform shape}, font=\Large},
						legend to name={legendcomparison1},
%						ymode=log,
%						log basis y={10},
						ymajorgrids=true,
						grid style=dashed,
						line width=3pt
						%scaled y ticks=false% crucial addition
						]
						\addplot[
						color=teal,
						mark=diamond,
						mark size=5pt,
						]
						coordinates {
							(4,7251)(8,6091)(12,5206)(16,4858)(20,3970)
						};
						\addplot[
						color=red,
						mark=triangle,
						mark size=5pt,
						]
						coordinates {
							(4,14201)(8,12982)(12,12216)(16,11281)(20,9430)
						};						
					\end{axis}
				\end{tikzpicture}
			}
			\captionsetup{justification=centering, font=scriptsize}
			\caption{\scriptsize Varying minSeason}
		\end{subfigure}
		\begin{subfigure}{0.32\columnwidth}
			\centering
			\resizebox{\linewidth}{!}{
				\begin{tikzpicture}[scale=0.6]
					\begin{axis}[
						compat=newest,
						xlabel={minDensity (\%)},
						ylabel={Memory Usage (MB)}, 
						label style={font=\Huge},
						ticklabel style = {font=\Huge},
						xmin=0.5, xmax=1.5,
						ymin=0, ymax=15000,
						xtick={0.5,0.75,1,1.25,1.5},
						ytick ={0, 5000,10000,15000},
						xticklabel style = {xshift=3mm}, %shift x tick label
						legend columns=-1,
						legend entries =  {DSTPM, APS},
						%legend image post style={scale=0.6},  %%% scale legend marker size
						legend style={nodes={scale=0.5,  transform shape}, font=\Large},
						legend to name={legendcomparison1},
						%ymode=log,
						%log basis y={10},
						ymajorgrids=true,
						grid style=dashed,
						line width=3pt
						%scaled y ticks=false% crucial addition
						]
						\addplot[
						color=teal,
						mark=diamond,
						mark size=5pt,
						]
						coordinates {
							(0.5,7091)(0.75,5500)(1,3106)(1.25,2274)(1.5,1842)
						};
						\addplot[
						color=red,
						mark=triangle,
						mark size=5pt,
						]
						coordinates {
							(0.5,12982)(0.75,11595)(1,9261)(1.25,7693)(1.5,7075)
							%(0.5,ABC)(0.75,ABC)(1,ABC)(1.25,ABC)(1.5,ABC)
						};						
					\end{axis}
				\end{tikzpicture}
			}
			\captionsetup{justification=centering, font=scriptsize}
			\caption{\scriptsize Varying minDensity}
		\end{subfigure}
		\begin{subfigure}{0.32\columnwidth}
			\centering
			\resizebox{\linewidth}{!}{
				\begin{tikzpicture}[scale=0.6]
					\begin{axis}[
						compat=newest,
						xlabel={maxPeriod (\%)},
						ylabel={Memory Usage (MB)}, 
						label style={font=\Huge},
						ticklabel style = {font=\Huge},
						xmin=0.2, xmax=1,
						ymin=0, ymax=20000,
						xtick={0.2,0.4,0.6,0.8,1},
						ytick ={0, 5000,10000,15000,20000},
						xticklabel style = {xshift=3mm}, %shift x tick label
						legend columns=-1,
						legend entries = {DSTPM, APS},
						%legend image post style={scale=0.6},  %%% scale legend marker size
						legend style={nodes={scale=0.5,   transform shape}, font=\Large},
						legend to name={legendcomparison1},
						%ymode=log,
						%log basis y={10},
						ymajorgrids=true,
						grid style=dashed,
						line width=3pt
						%scaled y ticks=false% crucial addition
						]
						\addplot[
						color=teal,
						mark=diamond,
						mark size=5pt,
						]
						coordinates {
							(0.2,6001)(0.4,6591)(0.6,7437)(0.8,8223)(1,9231)
						};
						\addplot[
						color=red,
						mark=triangle,
						mark size=5pt,
						]
						coordinates {
							(0.2,13074)(0.4,13982)(0.6,15631)(0.8,16992)(1,19934)
						};						
					\end{axis}
				\end{tikzpicture}
			}
			\captionsetup{justification=centering, font=scriptsize}
			\caption{\scriptsize Varying maxPeriod}
		\end{subfigure}
		%\vspace{-0.1in}
		%\ref{legendcomparison}
		\vspace{-0.08in}
		\captionsetup{justification=centering, font=small}
		\caption{Memory Usage Comparison on RE (real-world)}
		\label{fig:memorybaselineRE}
	\end{minipage}      
\end{figure*}
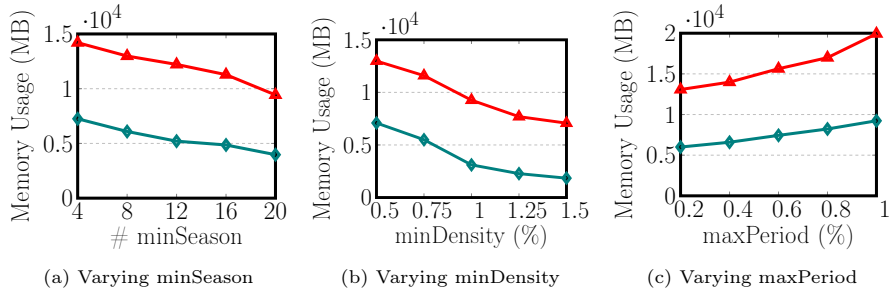  

\begin{figure*}[!t]
	\begin{minipage}[t]{1\columnwidth} 
		\centering
		\begin{subfigure}{0.32\columnwidth}
			\centering
			\resizebox{\linewidth}{!}{
				\begin{tikzpicture}[scale=0.6]
					\begin{axis}[
						compat=newest,
						xlabel={\# minSeason},
						ylabel={Memory Usage (MB)}, 
						label style={font=\Huge},
						ticklabel style = {font=\Huge},
						xmin=4, xmax=20,
						ymin=0, ymax=12000,
						xtick={4,8,12,16,20},
						ytick ={4000,8000,12000},
						legend columns=-1,
						legend entries =  {DSTPM, APS},
						legend style={nodes={scale=0.55,  transform shape}, font=\Large},
						legend to name={legendpruning},
						%ymode=log,
						%log basis y={10},
						ymajorgrids=true,
						grid style=dashed,
						line width=3pt,
						%scaled y ticks=false% crucial addition
						]
						\addplot[
						color=teal,
						mark=asterisk,
						mark size=5pt,
						]
						coordinates {
							(4,6012)(8,4321)(12,3547)(16,2981)(20,1401)
						};
						\addplot[
						color=red,
						mark=triangle,
						mark size=5pt,
						]
						coordinates {
							(4,10834)(8,9872)(12,8753)(16,8052)(20,6206)
						};						
					\end{axis}
				\end{tikzpicture}
			}
			\caption{\scriptsize Varying minSeason.}
		\end{subfigure}
		\begin{subfigure}{0.32\columnwidth}
			\centering
			\resizebox{\linewidth}{!}{
				\begin{tikzpicture}[scale=0.6]
					\begin{axis}[
						compat=newest,
						xlabel={minDensity (\%)},
						ylabel={Memory Usage (MB)}, 
						label style={font=\Huge},
						ticklabel style = {font=\Huge},
						xmin=0.5, xmax=1.5,
						ymin=0, ymax=12000,
						xtick={0.5,0.75,1,1.25,1.5},
						ytick ={4000,8000,12000},
						legend columns=-1,
						legend entries =  {DSTPM, APS},
						legend style={nodes={scale=0.55,  transform shape}, font=\Large},
						legend to name={legendpruning},
						%ymode=log,
						%log basis y={10},
						ymajorgrids=true,
						grid style=dashed,
						line width=3pt,
						%scaled y ticks=false% crucial addition
						]
						\addplot[
						color=teal,
						mark=asterisk,
						mark size=5pt,
						]
						coordinates {
							(0.5,5321)(0.75,3832)(1,3087)(1.25,2562)(1.5,2001)
						};
						\addplot[
						color=red,
						mark=triangle,
						mark size=5pt,
						]
						coordinates {
							(0.5,9872)(0.75,8183)(1,7046)(1.25,6749)(1.5,6201)
						};						
					\end{axis}
				\end{tikzpicture}
			}
			\caption{\scriptsize Varying minDensity.}
		\end{subfigure}
		\begin{subfigure}{0.32\columnwidth}
			\centering
			\resizebox{\linewidth}{!}{
				\begin{tikzpicture}[scale=0.6]
					\begin{axis}[
						compat=newest,
						xlabel={maxPeriod (\%)},
						ylabel={Memory Usage (MB)}, 
						label style={font=\Huge},
						ticklabel style = {font=\Huge},
						xmin=0.2, xmax=1,
						ymin=0, ymax=16000,
						xtick={0.2,0.4,0.6,0.8,1},
						ytick ={4000,8000,12000,16000},
						legend columns=-1,
						legend entries =  {DSTPM, APS},
						legend style={nodes={scale=0.55,   transform shape}, font=\Large},
						legend to name={legendpruning},
						%ymode=log,
						%log basis y={10},
						ymajorgrids=true,
						grid style=dashed,
						line width=3pt,
						%scaled y ticks=false% crucial addition
						]
						\addplot[
						color=teal,
						mark=asterisk,
						mark size=5pt,
						]
						coordinates {
							(0.2,4210)(0.4,4921)(0.6,6832)(0.8,7520)(1,8215)
						};
						\addplot[
						color=red,
						mark=triangle,
						mark size=5pt,
						]
						coordinates {
							(0.2,9274)(0.4,9872)(0.6,11134)(0.8,12853)(1,14531)
							%(0.2,ABC)(0.4,ABC)(0.6,ABC)(0.8,ABC)(1,ABC)
						};						
					\end{axis}
				\end{tikzpicture}
			}
			\caption{\scriptsize Varying maxPeriod.}
		\end{subfigure}
		%\vspace{-0.1in}
		%\vspace{-0.05in}
		\caption{Memory Usage Comparison on SC (real-world)}
		\label{fig:memorybaselineSC}
	\end{minipage}    
\end{figure*}
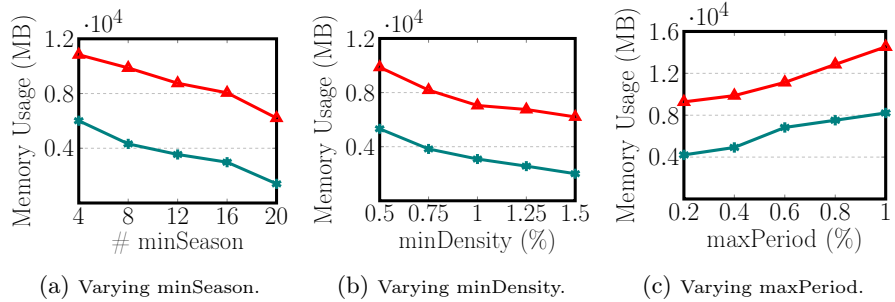 
\begin{figure*}[!t]
	\begin{minipage}{1\columnwidth}
		\centering
		\begin{subfigure}[t]{0.32\columnwidth}
			\centering
			\captionsetup{justification=centering}
			\resizebox{\linewidth}{!}{
				\begin{tikzpicture}[scale=0.6]
					\begin{axis}[
						xlabel={\# worker nodes},
						ylabel={Runtime (sec)}, 
						label style={font=\Huge},
						ticklabel style = {font=\Huge},
						xticklabel style = {xshift=3mm}, %shift x tick label
						xmin=1, xmax=20,
						ymin=0, ymax=300000,
						xtick={1,5,10,15,20},
						legend columns=-1,
						legend entries = {20 partitions, 40 partitions, 60 partitions},
						legend image post style={scale=0.6},  %%% scale legend marker size
						legend style={nodes={scale=0.5, transform shape}, font=\Large},
						legend to name={legendcomparison2},
						legend pos= north west,			
						ymajorgrids=true,
						grid style=dashed,
						ymode=log,
						log basis y={10},
						line width=3pt
						%scaled x ticks={base 10:-3},
						%every x tick label/.append style={alias=XTick,inner xsep=0pt},
						%every x tick scale label/.style={at=(XTick.base east),anchor=base west},
						%tick scale binop=\times
						]
						\addplot[
						color=blue,
						mark=x,
						mark size=5pt,
						]
						coordinates {
							(1,123610)(5,60245)(10,21004)(15,11654)(20,8562)
							%(1,ABC)(5,ABC)(10,ABC)(15,ABC)(20,ABC)
						};
						\addplot[
						color=teal,
						mark=square,
						mark size=5pt,
						]
						coordinates {
							(1,223610)(5,85245)(10,43504)(15,21654)(20,15562)
						};
						\addplot[
						color=red,
						mark=diamond,
						mark size=5pt,
						]
						coordinates {
							(1,291610)(5,115245)(10,59504)(15,27654)(20,19562)
						};
					\end{axis}
				\end{tikzpicture}
			}
			\captionsetup{justification=centering, font=scriptsize}
			\caption{RE}
			\label{fig:scaleRE1}
		\end{subfigure}
		\begin{subfigure}[t]{0.32\columnwidth}
			\centering
			\captionsetup{justification=centering}
			\resizebox{\linewidth}{!}{
				\begin{tikzpicture}[scale=0.6]
					\begin{axis}[
						xlabel={\# worker nodes},
						ylabel={Runtime (sec)}, 
						label style={font=\Huge},
						ticklabel style = {font=\Huge},
						xticklabel style = {xshift=3mm}, %shift x tick label
						xmin=1, xmax=20,
						ymin=0, ymax=300000,
						xtick={1,5,10,15,20},
						legend columns=-1,
						legend entries = {20 partitions, 40 partitions, 60 partitions},
						legend image post style={scale=0.6},  %%% scale legend marker size
						legend style={nodes={scale=0.5, transform shape}, font=\Large},
						legend to name={legendcomparison2},
						legend pos= north west,			
						ymajorgrids=true,
						grid style=dashed,
						ymode=log,
						log basis y={10},
						line width=3pt
						%scaled x ticks={base 10:-3},
						%every x tick label/.append style={alias=XTick,inner xsep=0pt},
						%every x tick scale label/.style={at=(XTick.base east),anchor=base west},
						%tick scale binop=\times
						]
						\addplot[
						color=blue,
						mark=x,
						mark size=5pt,
						]
						coordinates {
							(1,110610)(5,50245)(10,18004)(15,9654)(20,7562)
							%(1,ABC)(5,ABC)(10,ABC)(15,ABC)(20,ABC)
						};
						\addplot[
						color=teal,
						mark=square,
						mark size=5pt,
						]
						coordinates {
							(1,193610)(5,75245)(10,30504)(15,18654)(20,12562)
						};
						\addplot[
						color=red,
						mark=diamond,
						mark size=5pt,
						]
						coordinates {
							(1,251610)(5,95245)(10,49504)(15,25654)(20,17562)
						};
					\end{axis}
				\end{tikzpicture}
			}
			\captionsetup{justification=centering, font=scriptsize}
			\caption{SC}
			\label{fig:scaleSC1}
		\end{subfigure}
		\begin{subfigure}[t]{0.32\columnwidth}
			\centering
			\captionsetup{justification=centering}
			\resizebox{\linewidth}{!}{
				\begin{tikzpicture}[scale=0.6]
					\begin{axis}[
						xlabel={\# worker nodes},
						ylabel={Runtime (sec)}, 
						label style={font=\Huge},
						ticklabel style = {font=\Huge},
						xticklabel style = {xshift=3mm}, %shift x tick label
						xmin=1, xmax=20,
						ymin=0, ymax=300000,
						xtick={1,5,10,15,20},
						legend columns=-1,
						legend entries = {20 partitions, 40 partitions, 60 partitions},
						legend image post style={scale=0.6},  %%% scale legend marker size
						legend style={nodes={scale=0.5, transform shape}, font=\Large},
						legend to name={legendcomparison2},
						legend pos= north west,			
						ymajorgrids=true,
						grid style=dashed,
						ymode=log,
						log basis y={10},
						line width=3pt
						%scaled x ticks={base 10:-3},
						%every x tick label/.append style={alias=XTick,inner xsep=0pt},
						%every x tick scale label/.style={at=(XTick.base east),anchor=base west},
						%tick scale binop=\times
						]
						\addplot[
						color=blue,
						mark=x,
						mark size=5pt,
						]
						coordinates {
							(1,73610)(5,40245)(10,14004)(15,6154)(20,5562)
							%(1,ABC)(5,ABC)(10,ABC)(15,ABC)(20,ABC)
						};
						\addplot[
						color=teal,
						mark=square,
						mark size=5pt,
						]
						coordinates {
							(1,133610)(5,65245)(10,26504)(15,11654)(20,10562)
						};
						\addplot[
						color=red,
						mark=diamond,
						mark size=5pt,
						]
						coordinates {
							(1,191610)(5,85245)(10,39504)(15,14654)(20,13562)
						};
					\end{axis}
				\end{tikzpicture}
			}
			\captionsetup{justification=centering, font=scriptsize}
			\caption{INF}
			\label{fig:scaleINF1}
		\end{subfigure}
		%\vspace{-0.1in}
		\ref{legendcomparison2}
		%\vspace{-0.08in}
		\captionsetup{justification=centering, font=small}
		{\\ minSeason=4, minDensity=0.5\%, maxPeriod=0.2\%, \\\;distInterval=[90, 270] (RE, SC) and [30, 90] (INF)}
		\caption{Scalability evaluation} %Varying \% of sequences on RE}
		\label{fig:scale1}
	\end{minipage}
	\hspace{0.2in}
	\begin{minipage}{1\columnwidth}
		\centering
		\begin{subfigure}[t]{0.32\columnwidth}
			\centering
			\captionsetup{justification=centering}
			\resizebox{\linewidth}{!}{
				\begin{tikzpicture}[scale=0.6]
					\begin{axis}[
						xlabel={\# worker nodes},
						ylabel={Runtime (sec)}, 
						label style={font=\Huge},
						ticklabel style = {font=\Huge},
						xticklabel style = {xshift=3mm}, %shift x tick label
						xmin=1, xmax=20,
						ymin=0, ymax=300000,
						xtick={1,5,10,15,20},
						legend columns=-1,
						legend entries = {20 partitions, 40 partitions, 60 partitions},
						legend image post style={scale=0.6},  %%% scale legend marker size
						legend style={nodes={scale=0.5, transform shape}, font=\Large},
						legend to name={legendcomparison2},
						legend pos= north west,			
						ymajorgrids=true,
						grid style=dashed,
						ymode=log,
						log basis y={10},
						line width=3pt
						%scaled x ticks={base 10:-3},
						%every x tick label/.append style={alias=XTick,inner xsep=0pt},
						%every x tick scale label/.style={at=(XTick.base east),anchor=base west},
						%tick scale binop=\times
						]
						\addplot[
						color=blue,
						mark=x,
						mark size=5pt,
						]
						coordinates {
							(1,80610)(5,44245)(10,18004)(15,9654)(20,6562)
							%(1,ABC)(5,ABC)(10,ABC)(15,ABC)(20,ABC)
						};
						\addplot[
						color=teal,
						mark=square,
						mark size=5pt,
						]
						coordinates {
							(1,153610)(5,65245)(10,33504)(15,18654)(20,12562)
						};
						\addplot[
						color=red,
						mark=diamond,
						mark size=5pt,
						]
						coordinates {
							(1,191610)(5,85245)(10,40504)(15,26654)(20,17562)
						};
					\end{axis}
				\end{tikzpicture}
			}
			\captionsetup{justification=centering, font=scriptsize}
			\caption{RE}
			\label{fig:scaleRE2}
		\end{subfigure}
		\begin{subfigure}[t]{0.32\columnwidth}
			\centering
			\captionsetup{justification=centering}
			\resizebox{\linewidth}{!}{
				\begin{tikzpicture}[scale=0.6]
					\begin{axis}[
						xlabel={\# worker nodes},
						ylabel={Runtime (sec)}, 
						label style={font=\Huge},
						ticklabel style = {font=\Huge},
						xticklabel style = {xshift=3mm}, %shift x tick label
						xmin=1, xmax=20,
						ymin=0, ymax=300000,
						xtick={1,5,10,15,20},
						legend columns=-1,
						legend entries = {20 partitions, 40 partitions, 60 partitions},
						legend image post style={scale=0.6},  %%% scale legend marker size
						legend style={nodes={scale=0.5, transform shape}, font=\Large},
						legend to name={legendcomparison2},
						legend pos= north west,			
						ymajorgrids=true,
						grid style=dashed,
						ymode=log,
						log basis y={10},
						line width=3pt
						%scaled x ticks={base 10:-3},
						%every x tick label/.append style={alias=XTick,inner xsep=0pt},
						%every x tick scale label/.style={at=(XTick.base east),anchor=base west},
						%tick scale binop=\times
						]
						\addplot[
						color=blue,
						mark=x,
						mark size=5pt,
						]
						coordinates {
							(1,60610)(5,34245)(10,12004)(15,7654)(20,4562)
							%(1,ABC)(5,ABC)(10,ABC)(15,ABC)(20,ABC)
						};
						\addplot[
						color=teal,
						mark=square,
						mark size=5pt,
						]
						coordinates {
							(1,123610)(5,50245)(10,25504)(15,14654)(20,9562)
						};
						\addplot[
						color=red,
						mark=diamond,
						mark size=5pt,
						]
						coordinates {
							(1,151610)(5,62245)(10,34504)(15,20654)(20,12562)
						};
					\end{axis}
				\end{tikzpicture}
			}
			\captionsetup{justification=centering, font=scriptsize}
			\caption{SC}
			\label{fig:scaleSC2}
		\end{subfigure}
		\begin{subfigure}[t]{0.32\columnwidth}
			\centering
			\captionsetup{justification=centering}
			\resizebox{\linewidth}{!}{
				\begin{tikzpicture}[scale=0.6]
					\begin{axis}[
						xlabel={\# worker nodes},
						ylabel={Runtime (sec)}, 
						label style={font=\Huge},
						ticklabel style = {font=\Huge},
						xticklabel style = {xshift=3mm}, %shift x tick label
						xmin=1, xmax=20,
						ymin=0, ymax=300000,
						xtick={1,5,10,15,20},
						legend columns=-1,
						legend entries = {20 partitions, 40 partitions, 60 partitions},
						legend image post style={scale=0.6},  %%% scale legend marker size
						legend style={nodes={scale=0.5, transform shape}, font=\Large},
						legend to name={legendcomparison2},
						legend pos= north west,			
						ymajorgrids=true,
						grid style=dashed,
						ymode=log,
						log basis y={10},
						line width=3pt
						%scaled x ticks={base 10:-3},
						%every x tick label/.append style={alias=XTick,inner xsep=0pt},
						%every x tick scale label/.style={at=(XTick.base east),anchor=base west},
						%tick scale binop=\times
						]
						\addplot[
						color=blue,
						mark=x,
						mark size=5pt,
						]
						coordinates {
							(1,40610)(5,24245)(10,9004)(15,5654)(20,2562)
							%(1,ABC)(5,ABC)(10,ABC)(15,ABC)(20,ABC)
						};
						\addplot[
						color=teal,
						mark=square,
						mark size=5pt,
						]
						coordinates {
							(1,83610)(5,45245)(10,20504)(15,10654)(20,5562)
						};
						\addplot[
						color=red,
						mark=diamond,
						mark size=5pt,
						]
						coordinates {
							(1,101610)(5,52245)(10,34504)(15,14654)(20,7562)
						};
					\end{axis}
				\end{tikzpicture}
			}
			\captionsetup{justification=centering, font=scriptsize}
			\caption{INF}
			\label{fig:scaleINF2}
		\end{subfigure}
		%\vspace{-0.1in}
		\ref{legendcomparison2}
		%\vspace{-0.08in}
		\captionsetup{justification=centering, font=small}
		{\\minSeason=12, minDensity=1.0\%, maxPeriod=0.6\%, \\\;distInterval=[90, 270] (RE, SC) and [30, 90] (INF)}
		\caption{Scalability evaluation} %Varying \% of sequences on RE}
	\label{fig:scale2}
	\end{minipage}
\end{figure*}

Our method for distributed seasonal temporal pattern mining is referred to as DSTPM.  Since there are no existing methods performing distributed seasonal temporal pattern mining, we adapt the state-of-the-art method for recurring itemset pattern mining PS-growth \cite{kiran2019finding} to find seasonal temporal patterns, then compare performance (both runtime and memory usage) between DSTPM and the adaped PS-growth (APS).
The adapted  PS-growth is conducted through 2-phase process: (1) PS-growth is applied to find frequent recurring events, and (2), mine temporal patterns from extracted events. 
                                     
We compare DSTPM with the baseline in terms of the runtime and memory usage. This experiment is run on a single node in our cluster. Figs. \ref{fig:runtimebaselineRE}, \ref{fig:runtimebaselineSC}, \ref{fig:memorybaselineRE} and \ref{fig:memorybaselineSC} show the comparison on RE and SC datasets. The results on another dataset are reported in the technical report \cite{ho2025distributedseasonal}. Note that Figs. \ref{fig:runtimebaselineRE}-\ref{fig:memorybaselineSC} use the same legend.

As shown in Figs. \ref{fig:runtimebaselineRE} and \ref{fig:runtimebaselineSC}, DSTPM achieves the best runtime than the baseline even when it runs sequentially. On the tested datasets, the range and average speedups of DSTPM compared to the sequential baseline is $[3.7$-$8.5]$ and $5.2$ on average. 

In terms of memory consumption, as shown in Figs. \ref{fig:memorybaselineRE} and \ref{fig:memorybaselineSC}, DSTPM is more efficient than the baseline. The range and the average memory consumption of DSTPM compared to the baseline is: $[2.1$-$4.7]$ and $2.8$.
%\vspace{-0.1in}
\subsection{Scalability evaluation on synthetic datasets}\vspace{-0.02in}
To evaluate the scalability of DSTPM, from real-world datasets, we generate three synthetic datasets for RE, SC, and INF, each contains 1 million sequences and 5000 variables. We verify the scalability of DSTPM by fixing the data size, the \textit{minSeason}, \textit{minDensity}, \textit{maxPeriod}, and \textit{distInterval} thresholds, while changing the number of worker nodes (\# nodes = 1, 5, 10, 15, 20) and the number of partitions (\# partitions = 20, 40, 60).

Figs. \ref{fig:scale1} and \ref{fig:scale2} show the runtimes of DSTPM on the synthetic datasets under the distributed setting (y-axis in log scale). DSTPM scales very well in all configurations. Specifically, using 20-partitions always results in the best performance, compared to 40 and 60-partitions. This occurs because our cluster consists of 16 nodes; therefore, the 16-partition configuration makes the most efficient use of available resources while minimizing communication overhead. In contrast, using 64 partitions leads to the worst performance, which we attribute to the increased communication overhead caused by creating and distributing a larger number of tasks across nodes.

\section{Conclusion and Future Work}
This paper presents the first distributed algorithm for seasonal temporal pattern mining (DSTPM) that works well on large-scale datasets. DSTPM uses a distributed hierarchical lookup hash structure to enable fast computation. Moreover, the proposed measure \textit{maxSeason} enables efficient pruning of infrequent seasonal patterns, helping reduce the search space and minimize communication overhead. Experiments on real-world and synthetic datasets show that DSTPM outperforms the baseline and scales well to very large datasets. In future work, we plan to investigate approximate methods for DSTPM to further reduce computation by pruning unnecessary time series.

\bibliographystyle{splncs04}
\bibliography{references}
\end{document}